\documentclass[5p]{elsarticle}
\usepackage{lineno,hyperref}
\modulolinenumbers[5]
\usepackage{graphicx}
\usepackage{subcaption}
\usepackage{placeins}
\usepackage{multirow}
\usepackage{ulem}
\usepackage{soul}
\usepackage{xcolor}
\usepackage{mathptmx}
\usepackage{amsmath}
\usepackage{tabularx}

\biboptions{sort&compress}
\begin{document}
	
	\begin{frontmatter}
		
		\title{Development of a versatile modular system to study the ablation and plume expansion in a thin film }
		\author[affiliation1,affiliation2]{Renjith Kumar R\corref{correspondingauthor1}}
		\cortext[correspondingauthor1]{Email:renjithkumar.cpd@gmail.com}
		\author[affiliation1]{Pabitra Kumar Mishra}
		\author[affiliation1,affiliation2]{B R Geethika}
		\author[affiliation1]{Hem Chandra Joshi}
		\author[affiliation1,affiliation2]{Jinto Thomas\corref{correspondingauthor2}}
		\cortext[correspondingauthor2]{Email:jinto@ipr.res.in}
		\address[affiliation1]{Institute For Plasma Research, Bhat, Gandhinagar, Gujarat, 382428, India}
		\address[affiliation2]{Homi Bhabha National Institute, Training School Complex, Anushaktinagar, Mumbai, 400094, India}
		
		\begin{abstract}
			
			Time-resolved investigation of laser ablation of thin films requires stable, repeatable, and efficient experimental environment. Conventional approaches often rely on multiple laser-ablation shots acquired at different delay times to reconstruct the temporal evolution of the process, thereby assuming shot-to-shot reproducibility. 
		This work presents a novel modular experimental setup that addresses these challenges and enables a systematic pump--probe study of laser--matter interaction. The setup provides reliable single-pulse measurements with controlled sample positioning inside the high vacuum chamber, as well as integrated data recording, which improves redundancy and ease with which experiments can be performed. Further, the setup serves as an advanced diagnostic tool for investigating the time evolution of laser--matter interactions, ablation dynamics and laser-produced plasmas in a single laser shot, in various configurations with a temporal resolution better than 15 ns. 
			Using this capability, we show that for a fixed ablation geometry, the transient reflectivity and transmission response depends strongly on the probing side, indicating that probing direction is an independent diagnostic variable whose omission can lead to incomplete or misleading picture of the ablation dynamics. This experimental technique can advance the understanding of surface modification and plume expansion, which are relevant to fundamental studies and also for range of applications, including laser cleaning of tokamak viewports and mirrors, pulsed laser deposition (PLD), surface structuring and laser-induced forward transfer (LIFT), among others. The present setup is versatile enough to perform these studies under different ablation configurations without implementing major changes in the experimental configuration.
		\end{abstract}
		\begin{keyword}
			Laser ablation, instrumentation, vacuum chamber, automation, thin films, pump--probe
			
		\end{keyword}
		
	\end{frontmatter}
	
	\section{Introduction}
	
	Laser--matter interaction in thin films has attracted substantial attention because of its relevance to surface modification, micro-machining, laser scribing, laser induced forward transfer(LIFT), thin-film processing, laser cleaning, pulsed laser deposition, and time-resolved diagnostics of transient material response\cite{chichkov1996femtosecond, thomas2023review, radziemski2020lasers, kwok2016developments,Vorobyev,kuk_laser_scribing,ishteev_laser_scribing,Mattle_LIFT,veiko_LIFT,Narihara_blow_off}. Furthermore, laser ablation of thin films provides a versatile platform for studying the fundamental physics of laser–matter interaction, such as plasma formation and evolution, inverse bremsstrahlung absorption, laser-supported detonation waves, plasma shielding effects, and material redeposition processes\cite{zhang_ns_fs_laser_ablation, brown_ablation,ashfold_PLD, renjith_jolt}. It is important to note that thin film ablation studies offer the possibility of probing the ablation process from both the front and rear sides, a versatility that no other laser ablation geometries can claim. Also, in thin-film laser ablation, material removal can occur in two distinct geometries: front-side ablation (FA) and rear-side ablation (RA). In front-side ablation, the laser beam is incident directly on the film surface, whereas in rear-side ablation, the laser irradiates the film through the transparent substrate, resulting in the ablation at the film–substrate interface \cite{Kim_FA_RA,Canteli_FA_RA, Chen_FA_RA, hallum_lift_off,schultze1991blow}. The underlying ablation mechanisms in these two configurations are fundamentally different. In RA, material removal predominantly occurs through a laser-induced lift-off process at the film–substrate interface, while FA generally leads to comparatively lower material removal under similar irradiation conditions \cite{Kim_FA_RA,singh_blow_off,George_blow_off}.
	
	To obtain a deeper understanding of front and rear ablation and material removal process, different diagnostic techniques have been employed, including time-resolved pump–probe imaging, spectroscopic diagnostics of the ablation plasma, and post-ablation surface characterization methods \cite{harilal2022optical,mondal2019neutral,bonse_optical_probing}. In addition, time-resolved reflectivity measurements have been widely used by several groups to investigate the phase transitions occurring during nanosecond laser ablation\cite{boneberg2000nanosecond,carrasco2015visualization}. Furthermore, thin films are particularly advantageous for transmission-based investigations because of their minimal material volume, which enables improved probing of the absorption and scattering processes, as well as the surface modifications occurring during the ablation process \cite{renjith_jolt,renjith_fa_p1}.

	Pump--probe studies have investigated thin-film systems using different excitation and probing geometries. Mahdieh et.al \cite{Mahdieh2012Optical,Mahdieh2013Experimental} investigated thin-film
	laser ablation using front-side ablation with front- and rear-side probing,
	as well as CW-probe transmission during ns ablation; however, rear-side
	ablation and complete set of ablation--probing configurations were not
	addressed. Moreover, the evolution and dynamics is investigated using detectors that are slow, resulting in a completely skewed response resulting the loss of information during the melting phase. Miao et al. \cite{Miao2014Study} and Yan et al. \cite{Yan2017Long}
	demonstrated two and four pump--probe-side configurations, respectively, but
	were confined to thermoreflectance studies of thermal transport and stress rather than laser ablation, with no transmission measurement. Rapp et al.
	\cite{Rapp2016Ultrafast} and Pflug et al. \cite{Pflug2018Case} developed
	ultrafast pump--probe reflectometry/ellipsometry systems for laser--matter
	interaction, however, without the complete combination of ablation/probing
	geometries and transmission measurements. Pflug et al.
	\cite{Pflug2024Transient} demonstrated front- and rear-side probing
	during thin-film ablation, while our previous work \cite{renjith_jolt} focused on rear-side ablation with ambient-pressure
	effects without systematically varying the probing direction. These studies
	indicated the need of a unified approach for systematic investigation of thin-film ablation using different ablation and probing geometries, simultaneous reflection and transmission measurements, and controlled ambient conditions.
	
	Although these studies have established important individual aspects of pump--probe diagnostics, systematic investigation combining the ablation and probing geometries in thin-film laser ablation still remains limited. Moreover, conventional time-resolved pump--probe measurements generally reconstruct the temporal evolution by performing separate measurements at different pump--probe delays, thereby relying on shot-to-shot reproducibility of the ablation event\cite{harilal2022optical}. Such an approach can be affected by pulse-to-pulse variations
	in laser energy and by changes in the ablation site. A single-shot measurement
	of the complete transient, on the other hand, can provide a more direct representation of the evolution of an individual ablation event.
	
	In the present study, we employ a continuous-wave (CW) 532~nm probe laser in conjunction with high-gain avalanche photodetectors to record the transient reflection and transmission during a single laser-ablation event. The modular platform enables FA--FP, FA--RP, RA--FP, and RA--RP configurations without major optical realignment, while reflection and transmission are acquired
	simultaneously. The systems further integrates single-pulse selection,
	automated sample positioning, synchronized data acquisition, controlled
	ambient-pressure operation, and automated data management and processing.
	This combination enables reproducible measurements while avoiding the need
	to reconstruct the individual transient from multiple ablation shots at
	different pump--probe delays.

	\section{Experimental Setup}
	
	\subsection{ Mechanical Design of experimental chamber}

\begin{figure}[htbp]
		\centering
		
		\begin{subfigure}[t]{0.48\columnwidth}
			\centering
			\includegraphics[width=\linewidth]{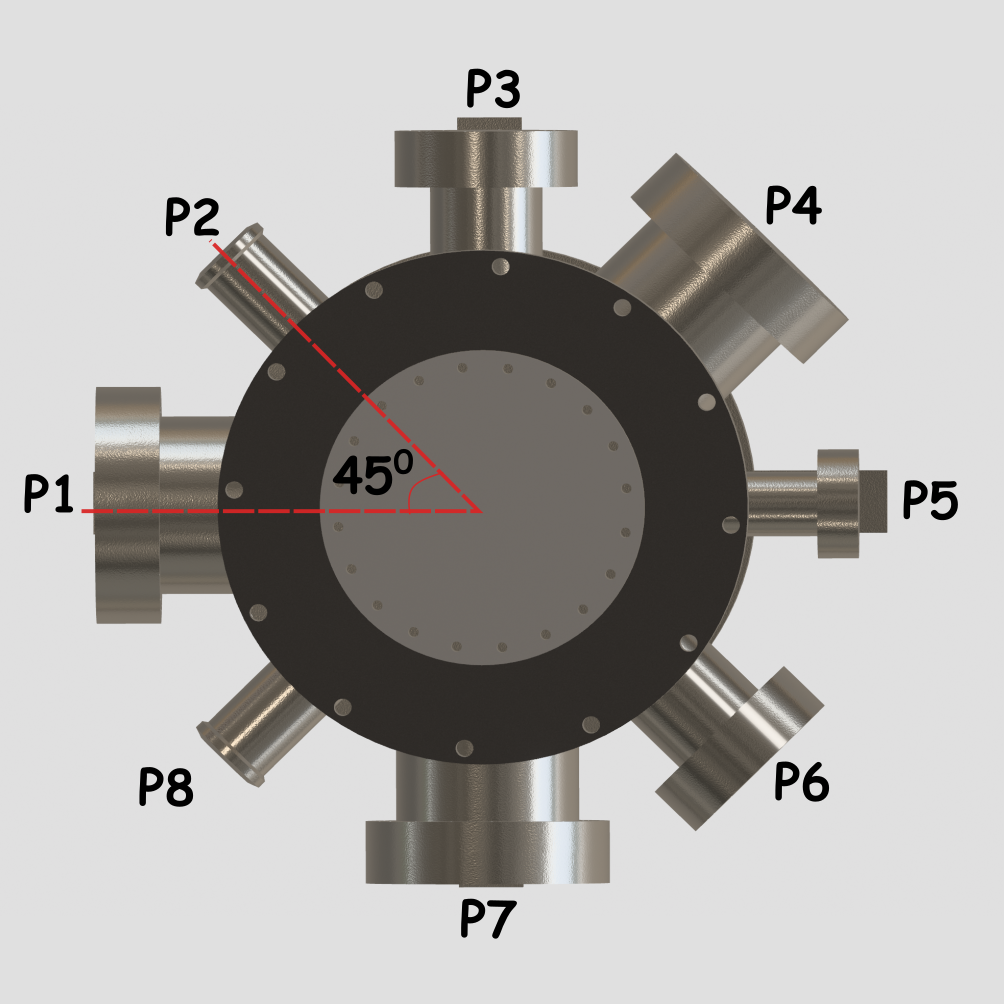}
			\caption{Top view and ports}
			\label{fig:Chamber_topview}
		\end{subfigure}
		\hfill
		\begin{subfigure}[t]{0.48\columnwidth}
			\centering
			\includegraphics[width=\linewidth]{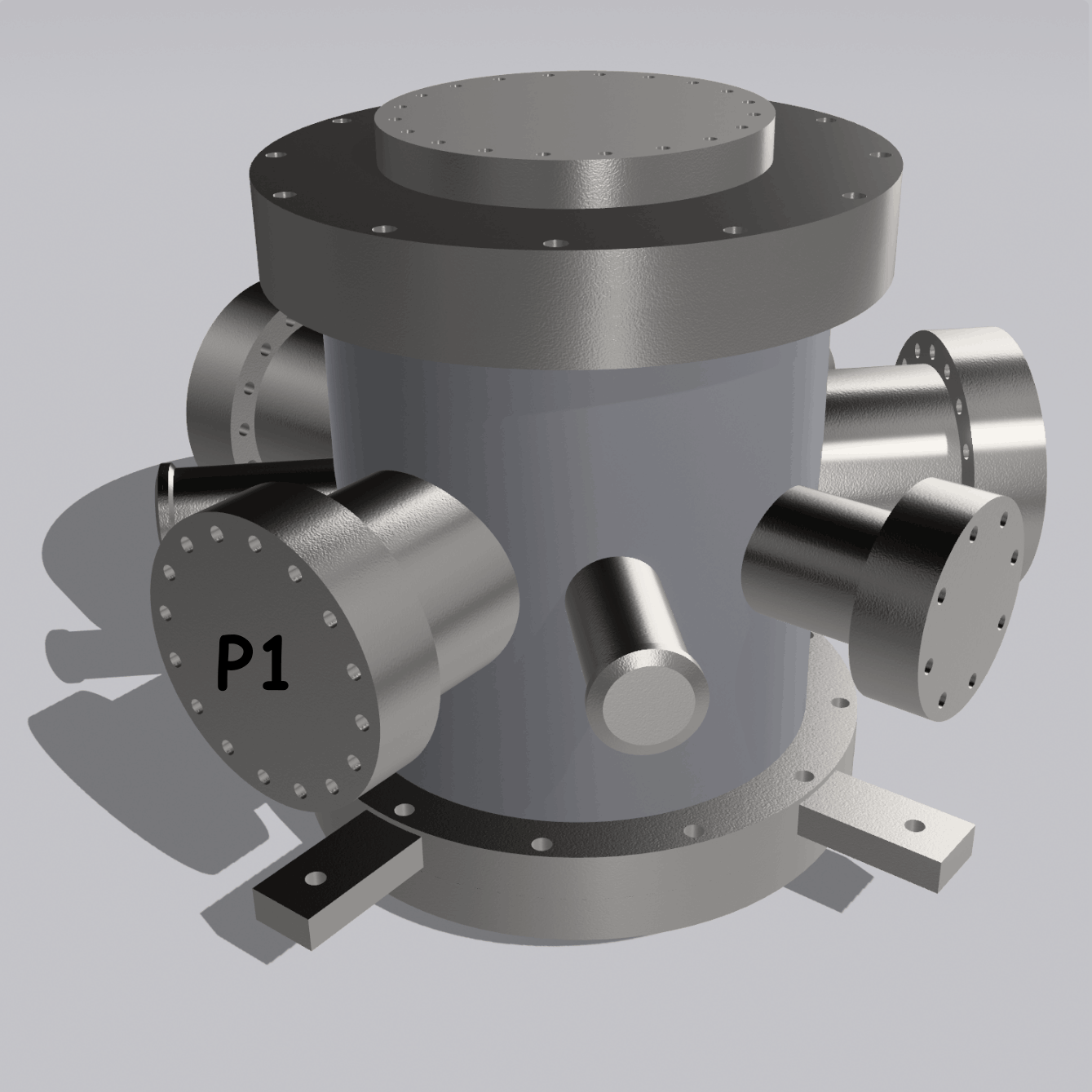}
			\caption{Isometric view}
			\label{fig:Chamber_isometric_view}
		\end{subfigure}
		
		\caption{Schematic view of the custom-designed vacuum chamber showing the top and isometric views with multiple radially distributed ports(P1 to P8). Items were summerised on table \ref{tab:components_list}}
		\label{fig:chamber_CAD_image}
	\end{figure}

\begin{figure*}[hbtp]
		\centering
		\includegraphics[width=0.75\textwidth]{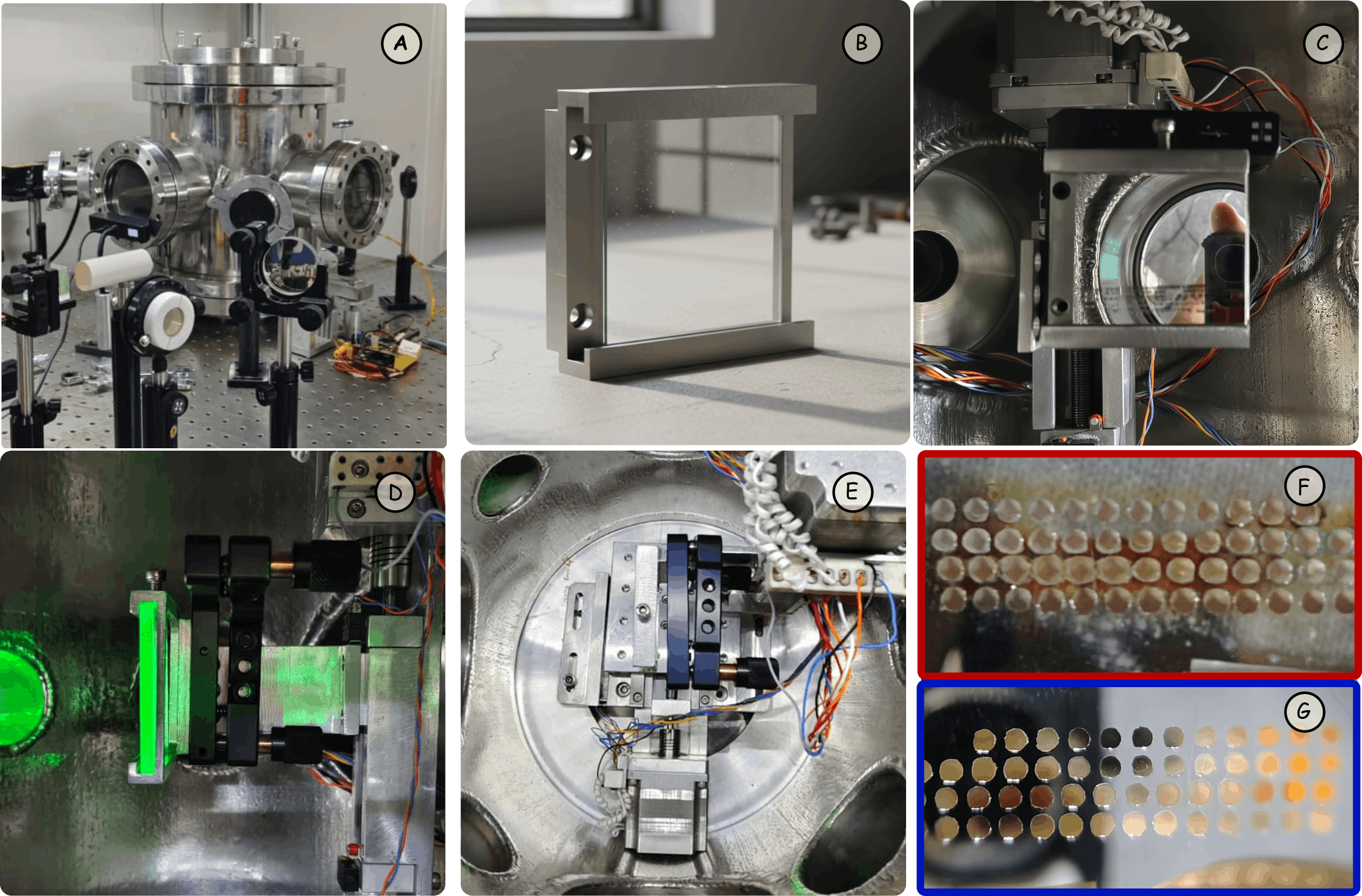}
		\caption{
			Experimental setup images 
			(A) Overview of the vacuum chamber integrated with external optical components. 
			(B) Sample holder with glass of 50x50x4mm mounted on it. 
			(C) Front view of sample mounted inside the chamber. 
			(D) Side view of sample mounted inside the chamber. 
			(E) Top view of sample mounted inside the chamber showing X and Z translation stage. 
			(F) Ablation spots after front ablation on Al 400nm thin film from front side of the sample
			(G) Ablation spots after front ablation on Al 400nm thin film from rear side of the sample
		}
		\label{fig:Exp_sys_pic}
	\end{figure*}

	Figure \ref{fig:chamber_CAD_image} shows the mechanical design of experimental vacuum chamber used including top view in figure \ref{fig:Chamber_topview} and isometric view in figure \ref{fig:Chamber_isometric_view}. The chamber is made of Stainless steel (SS-316) and has an inner diameter of 270~mm and a height of 266~mm and wall thickness of 9~mm. The ports of this chamber are strategically positioned to support the required optical, diagnostics alignments and to suit the evacuation and pressure measurements. Three 100~CF ports were incorporated: one for observation and sample replacement, other for pump-laser input and camera monitoring, and the third for the vacuum pump line, Pirani gauge, and gas feed-through connection using suitable reducers. Two 40~KF viewports were provided for probe-laser input and reflected-signal collection. In addition, two 63~CF ports were included, one for collecting the transmitted probe beam and the other for electrical feed-through connections. A 35~CF port was also provided for pump-laser beam delivery. Ports and usage are summerised Table~\ref{tab:components_list}. The Figure \ref{fig:Exp_sys_pic}A shows the fabricated experimental chamber.  Table~\ref{tab:components_list} provides the list of ports available on the experimental system and its angle of orientation with respect to the laser entrance port for ablation(P1)  The chamber on fabrication is tested for its leak tightness and observed to have a leak rate of 2$\times$10$^{-7}$~mbarL/s , which is sufficiently high enough to attain a vacuum of $10^{-6}$~mbar, for carrying out the laser plasma experiments.
	
	The chamber provides a mechanically rigid and modular structure, allowing convenient placement of optical ports, sample holders, and auxiliary components. This is particularly advantageous for pump--probe experiments, where stable optical alignment and sufficient diagnostic access are essential during long measurement runs. The flexible port configuration also provides scope for integrating additional diagnostics in future extensions.
	
	\begin{table*}[t]
		\centering

		\begin{tabular}{c l c l}
			\hline
			\textbf{Port} & \textbf{Component} & \textbf{Angle from P1 (°)} & \textbf{Purpose} \\
			\hline
			P1 & 100 CF viewport & 0   & Pump laser beam input and target observation \\
			P2 & 40 KF viewport & 45  & Probe laser beam entrance\\
			P3 & 63 CF electrical feedthrough & 90 & Electrical connections for translation stage \\
			P4 & 100 CF port & 135 & Vacuum pump and pressure gauge connection \\
			P5 & 35 CF viewport & 180 & Pump laser beam exit \\
			P6 & 63 CF viewport & 225 & Probe beam transmission detection \\
			P7 & 100 CF viewport & 270 & Sample observation and loading/unloading \\
			P8 & 40 KF viewport & 315 & Probe beam reflection detection \\
			\hline
		\end{tabular}
		\footnotesize
		\caption{Ports and their purpose in the chamber. All the viewports are made of fused silica.}
		\label{tab:components_list}
	\end{table*}
	
\subsection{Ablation and probing geometries}

\begin{figure}[hbtp]
		\begin{minipage}{0.45\textwidth}
			\centering
			\includegraphics[width=1\textwidth]{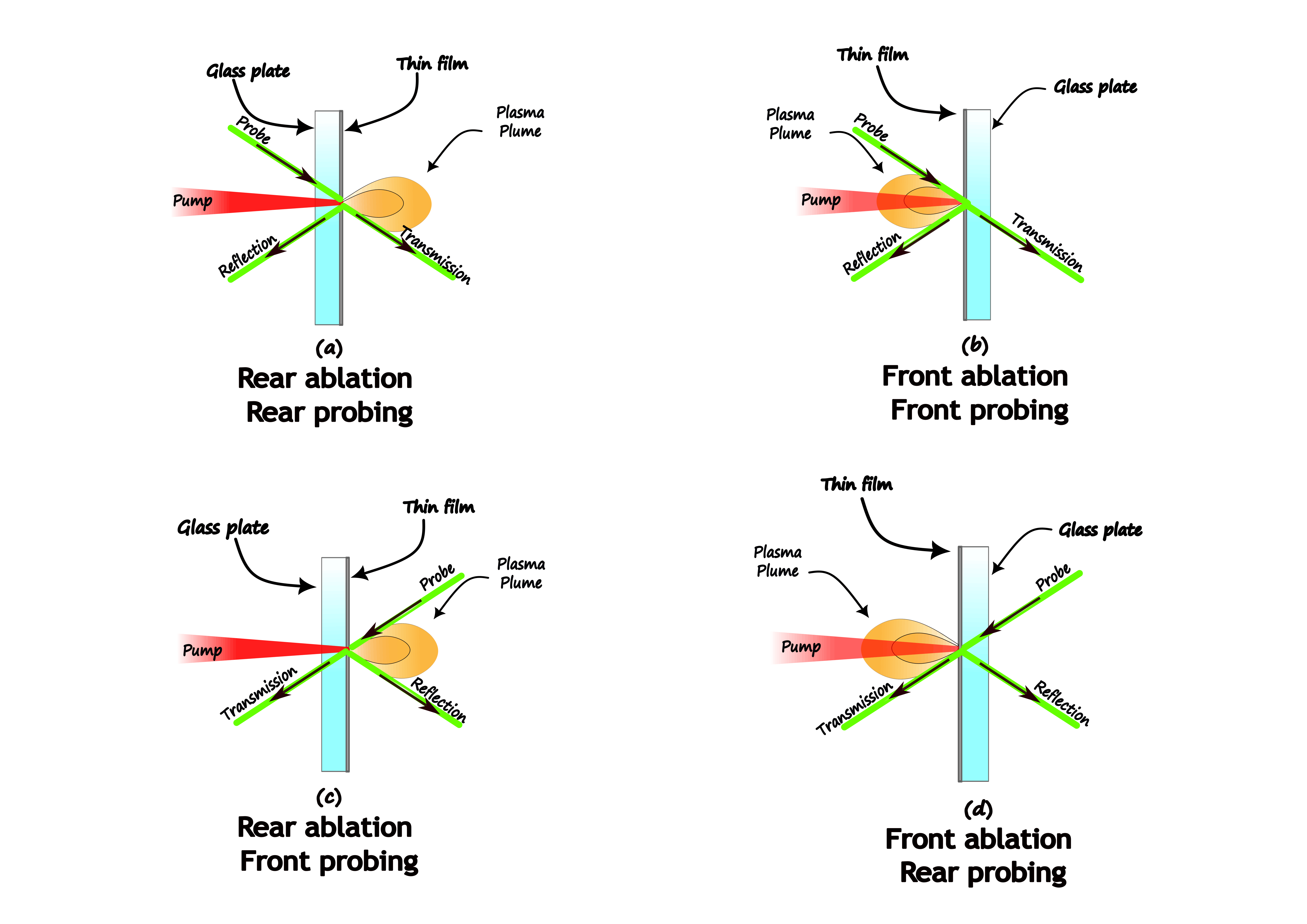}
			\caption{Different probing and ablation geometries of thin film }
			\label{fig:Different_orientation_for_pump_probe}
		\end{minipage}
	\end{figure}

	In the conventional laser ablation of bulk material, the laser is always incident directly on the sample surface and the plasma plume expands against the laser direction. However, when the ablation is carried out on a thin film coated on a transparent substrate the ablation modes can be investigated more comprehensively by varying the probing direction as well as ablation configuration as FA and BA, as illustrated in Fig. \ref{fig:Different_orientation_for_pump_probe}. In this scheme, probing from the film side corresponds to front probing, whereas probing film through the substrate corresponds to rear probing. Accordingly, four ablation–diagnostic configurations are possible: front ablation with front probing, front ablation with rear probing, rear ablation with front probing, and rear ablation with rear probing. 
		
	\subsection{High Vacuum compatible two directional sample manipulator}
	A high vacuum-compatible $X$--$Z$ motorized translation stage with a 50~mm travel range in either direction orthogonal to the pump laser beam is conceptualized to position the thin film sample within the available space and integrated inside the chamber. The sample manipulator is strategically placed inside the chamber to avoid the hindrances to of line of sight  of from different viewports for diagnostics.  The sample holder is fabricated from aluminium and consists of a metallic frame with a central opening for optical access. The thin-film-coated glass substrate is inserted into the frame and secured from the top using an M4 screw.
	The holder is designed to accommodate samples with dimensions of $50 \times 50 \times 5$~mm and is shown in Fig.~\ref{fig:Exp_sys_pic}(B).  The sample holder is coupled to the translation stage with a vacuum compatible kinematic mount to accommodate any minor variation in tilt and pitch while mounting the sample to precisely positioning the target. 
	
	Figure \ref{fig:Exp_sys_pic}(C) shows the front view of the sample mounted on the custom-designed holder, which is fixed to a kinematic mount and coupled to a motorized linear translation stage. The configuration ensures that the coated film surface is fully exposed, allowing maximum utilization of the sample area for laser ablation experiments. Figure \ref{fig:Exp_sys_pic}(D) presents the side view of the mounted assembly, where the adjustment knobs of the kinematic mount are clearly visible. Their placement ensures that they do not obstruct the probe laser beam, thereby maintaining an unobstructed optical path for pump--probe measurements. Figure \ref{fig:Exp_sys_pic}(E) shows the top view of the mounted sample, highlighting the optical accessibility available for the pump--probe configuration. Figures \ref{fig:Exp_sys_pic}(F) and \ref{fig:Exp_sys_pic}(G) present the laser-induced ablation spots on a 400 nm aluminium thin film coated on a quartz substrate of dimensions 50x50x4mm  following rear-side and front-side irradiation, respectively. The uniformly spaced and well-defined ablation craters demonstrate the high positioning accuracy and excellent shot-to-shot reproducibility achieved by the motorized translation stage. Furthermore, the automated translation enables systematic exposure of a fresh region of the film after each laser pulse, significantly reducing the time required for repeated measurements while maximizing the usable sample area. All mechanical components were carefully selected and integrated to ensure compatibility with the vacuum environment and to satisfy the optical constraints of the pump--probe experimental configuration.

	\subsection{Pump laser and its alignment}
\begin{figure}[hbtp]
 		\begin{minipage}{0.45\textwidth}
 			\centering
 			\includegraphics[width=1\textwidth]{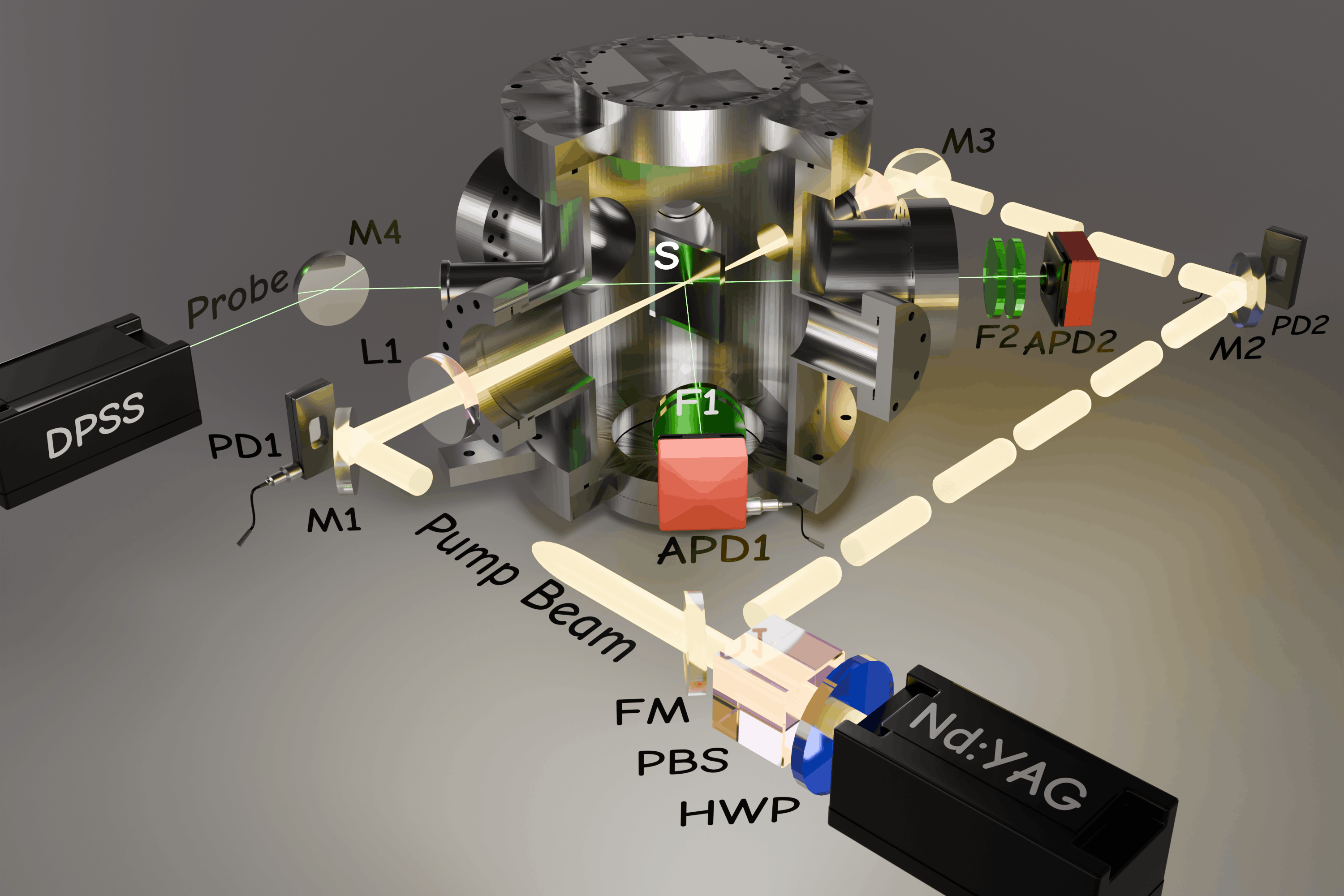}
 			\caption{Experimental pump probe setup for measuring transmittance and reflectance. DPSS-probe laser, Nd:YAG- pump laser, M-Mirror, FM- flip mirror for path change if need, PD- Photodiode, F-interference filter, APD- Avalanche photodiode, PBS-Polarising beam splitter, HWP-Half wave plate, S- sample.}
 			\label{fig:optical_arrangements}
 		\end{minipage}
 	\end{figure}

 	As discussed above, thin-film ablation may be carried out in two configurations, i.e., front-side and rear-side ablation. These two geometries are fundamentally different with respect to plasma parameters, ablation mechanism, etc \cite{Kim_FA_RA,Canteli_FA_RA, Chen_FA_RA, hallum_lift_off,schultze1991blow,Mondal_FA_RA,renjith_jolt}. Therefore, probing can also be arranged in either front or rear geometry. The possible pump--probe orientations are summarized in Fig.~\ref{fig:Different_orientation_for_pump_probe}. Figure~\ref{fig:optical_arrangements} illustrates the complete optical arrangement employed for front- and rear-side laser ablation and pump--probe measurements. Laser ablation was performed using the fundamental wavelength (1064~nm) of a Q-switched Nd:YAG laser with a pulse duration of approximately 10~ns with 30Hz repetition rate. The pulse energy was continuously adjusted using a half-wave plate (HWP) in combination with a polarizing beam splitter (PBS), providing precise control of the laser fluence incident on the sample. The pump beam was subsequently directed to either the front- or rear-side irradiation path using a flip mirror(FM) though port P1 and P5, allowing switching between the two ablation geometries without disturbing the optical alignment. The selected beam was then focused onto the sample through either the 100 CF or the 35 CF viewport using a plano-convex lens, thereby enabling front-side or rear-side ablation, respectively. This arrangement ensured reproducible beam alignment while minimizing the time required to alternate between the two experimental configurations.
 		
\begin{figure}[hbtp]
		\begin{minipage}{0.45\textwidth}
			\centering
			\includegraphics[width=1\textwidth]{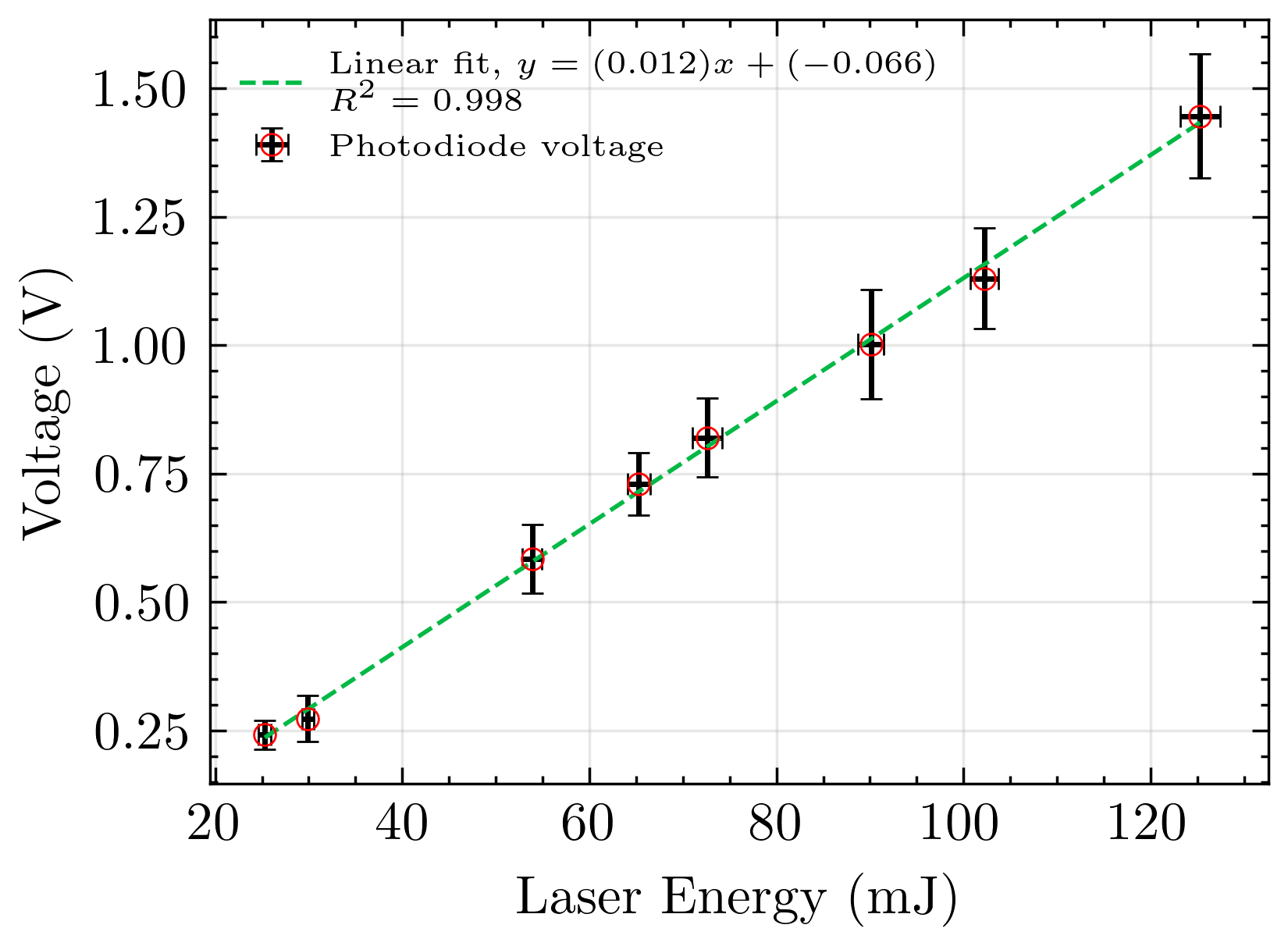}
			\caption{Energy calibration curve for the Photo diode, output voltage of photo diode versus pump laser energy. }
			\label{fig:energy_cal}
		\end{minipage}
	\end{figure}

 	The pulse energy of the Nd:YAG laser was continuously monitored using a fast photodiode (EOT - ET-3000) positioned adjacent to the laser output in front and rear ablation paths. Prior to the experiments, the photodiode response was calibrated against a commercial laser energy meter (OPHIR Laser Star meter and PE50 head), with the laser pulse energy independently measured over the operating energy range. Figure~\ref{fig:energy_cal} shows the resulting calibration curve, which exhibits a linear response. The calibration equation obtained from the linear fit was subsequently incorporated into the data acquisition software to convert the measured photodiode voltage into the corresponding pulse energy on shot-to-shot basis. This calibration enabled continuous monitoring of the pump laser energy and facilitated normalization of the transient signals to compensate for pulse-to-pulse energy fluctuations. A rotary solenoid (MSM GDAY 050 X20 E22) is employed as the shutter actuator for the pump laser and is driven using a ULN2003 driver module. 
	
	 \subsection{Probe laser }
	A continuous-wave diode-pumped solid-state (DPSS) laser operating at 532~nm with an output power of approximately 100~mW with a spot size of $\approx$~1mm was employed as the probe beam. The probe laser entered the vacuum chamber through the 40 KF viewport P2. The probe beam was incident at an angle of $45^\circ$
	with respect to the sample normal to facilitate simultaneous measurement of the reflected and transmitted signals. A series of precision apertures with diameter $\approx$4mm were positioned along the optical path before and after the chamber to define the beam trajectory and ensure accurate alignment through the sample. The reflected probe beam exited through the 40 KF viewport P8 and was collected by a second photodetector APD1 to measure the transient reflectivity. The transmitted probe beam exited the chamber through the opposite 63 CF viewport P6 and being directed onto a photodetector APD2 for transmission measurements.  The optical arrangement was configured such that the probe beam alignment remained fixed throughout the experiments, while front-side and rear-side probing geometries were changed by rotating the sample holder by $180^\circ$ This configuration ensured reproducible beam alignment, efficient collection of both reflected and transmitted signals, and reliable pump--probe measurements under identical experimental conditions.
	
	\subsection{Detectors for probe laser measurement}
	
\begin{figure}[hbtp]
		\begin{minipage}{0.45\textwidth}
			\centering
			\includegraphics[width=1\textwidth]{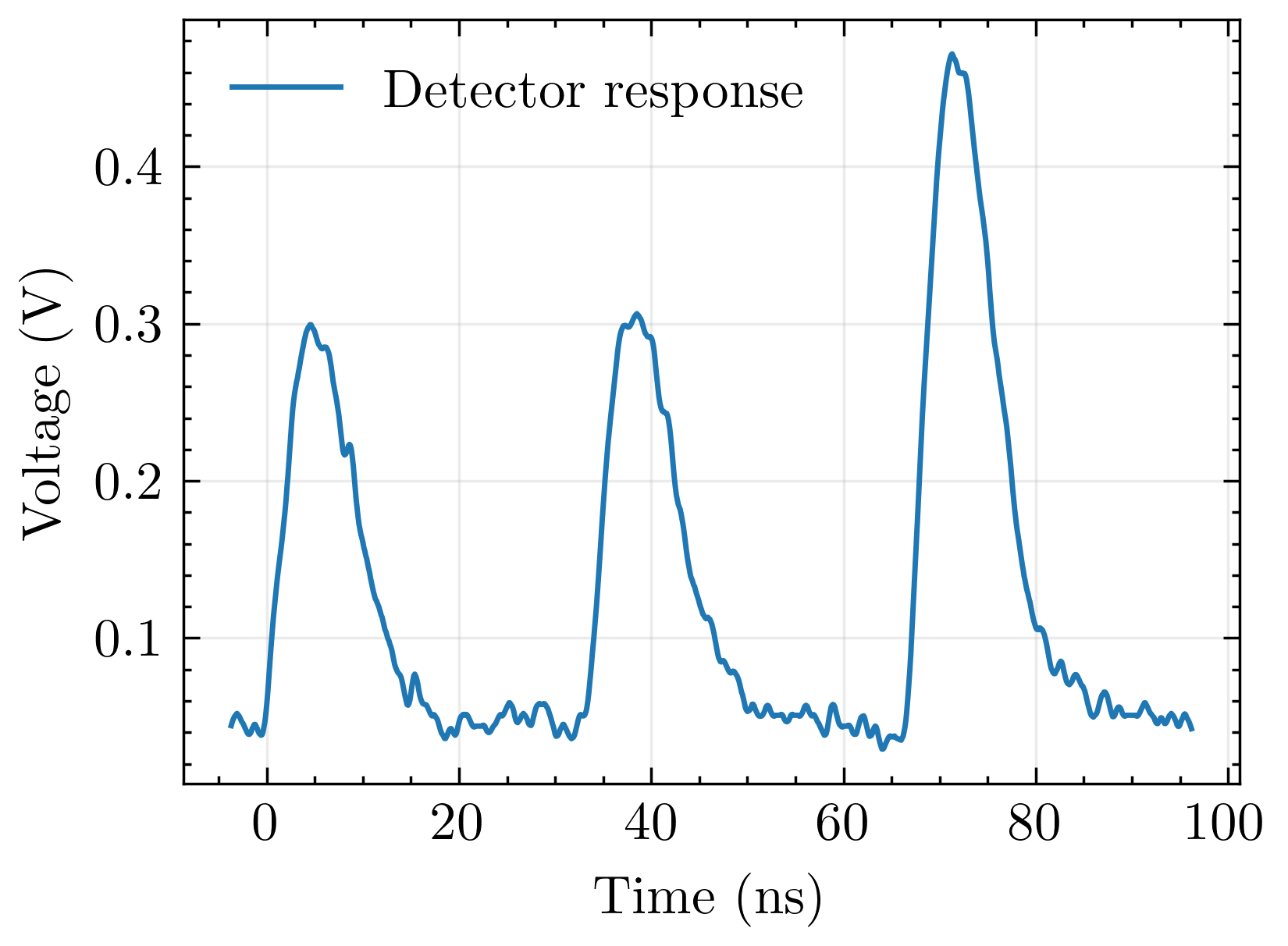}
			\caption{Temporal response of APD used for the measurement to a 30 M Hz, 0.3 ns pulsed train of super continuum source. }
			\label{fig:detector_response}
		\end{minipage}
	\end{figure}

	The reflected and transmitted probe beams were collected independently using two high-gain avalanche photodiodes (Hamamatsu S8890-02), positioned at the corresponding reflection and transmission ports of the vacuum chamber. Since the pump laser intensity is approximately nine orders of magnitude higher than that of the probe laser, effective suppression of scattered pump radiation is essential for reliable detection of the weak probe signal. Therefore, a combination of a 532 nm band-pass filter and a 650 nm short-pass filter was placed in front of each avalanche photodiode (APD), ensuring efficient transmission of the probe beam while rejecting the 1064 nm pump radiation and a significant fraction of the broadband plasma emission. Furthermore, the APDs were positioned after the condenser lenses, which collected the transmitted or reflected probe light and focused it onto the detector active area. This arrangement maximized the collected optical signal while minimizing sensitivity to slight beam divergence, mechanical vibrations, and minor alignment fluctuations during the experiment.  An amplifier circuit is connected to it to get the high gain detection.The combined detector-electronics system exhibited an effective response time of approximately 5--10 ns, which is sufficient to resolve the transient optical signals associated with nanosecond laser ablation. The temporal response of the complete detection system was characterized using a pulsed supercontinuum white-light source (NKT Photonics SuperK EVO) with a pulse width of approximately 0.3 ns and 30~MHz repetition rate, as shown in Fig.~\ref{fig:detector_response}. The recorded pulse exhibits a full width at half maximum (FWHM) of approximately 10 ns, indicating an overall temporal response of the detection system of about 10 ns.

	In addition, the detector electronics and signal cables were carefully shielded and grounded to minimize electromagnetic interference and pump-laser-induced electrical pickup. A fast photodiode positioned adjacent to the pump laser provided the trigger signal for the digital oscilloscope (MSO 7054B), establishing a common temporal reference for all measurements. This combination of high-speed detectors, custom electronics, optical filtering, and synchronized triggering enabled stable, high signal-to-noise ratio measurements with excellent shot-to-shot reproducibility over the entire duration of the pump--probe experiments.

	\subsection{Motion control and sample positioning}
	
\begin{figure}[hbtp]
		\begin{minipage}{0.45\textwidth}
			\centering
			\includegraphics[width=1\textwidth]{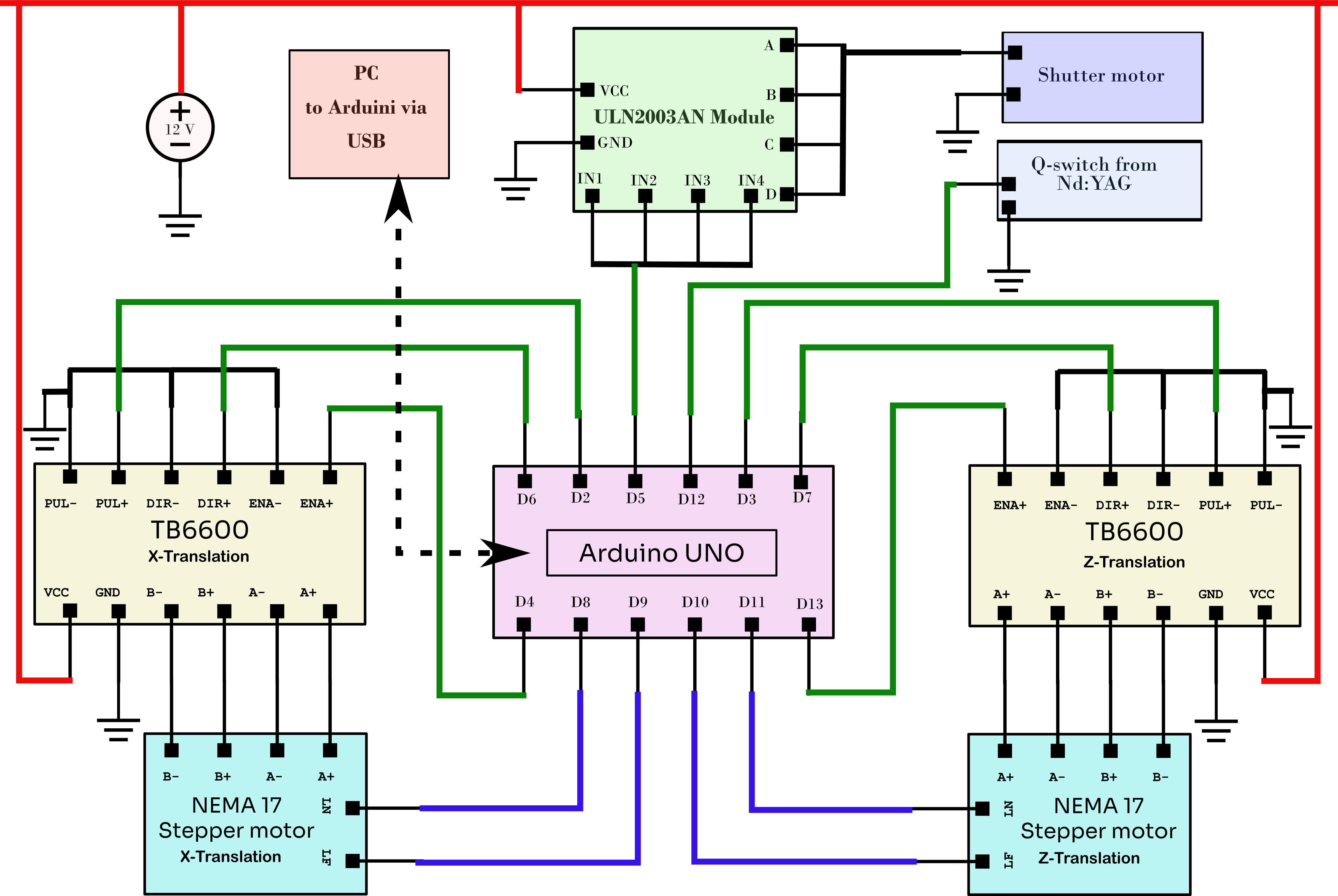}
			\caption{Electrical connection diagram of the control system showing Arduino UNO interfacing with TB6600 drivers, NEMA 17 motors, and ULN2003-based shutter control. A 12~V supply powers the system.}
			\label{fig:Exp_sys_connection_diagram}
		\end{minipage}
	\end{figure}

\begin{figure}[hbtp]
		\begin{minipage}{0.4\textwidth}
			\centering
			\includegraphics[width=1\textwidth]{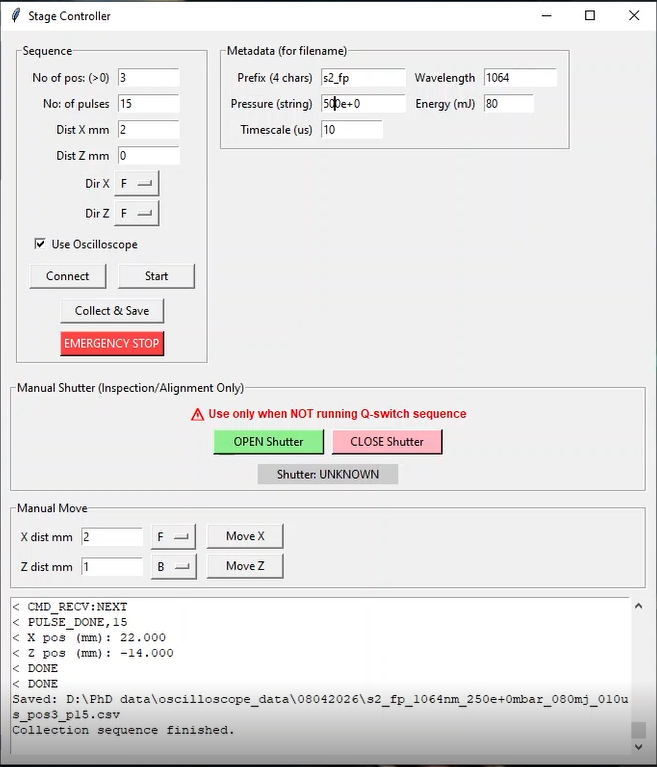}
			\caption{GUI developed for conducting the melt dynamics study of thin film. The GUI is used to configure the experiments which decides the number of spots and number of pulses in every spots, movement of sample along with laser parameters for the experiment.}
			\label{fig:GUI}
		\end{minipage}
	\end{figure}

	The sample positioning system employs motorized X and Z translation stages driven through TB6600 stepper motor drivers. The electronic connections of the control system are shown in Fig.~\ref{fig:Exp_sys_connection_diagram}. A NEMA~17 stepper motor, driven by a TB6600 driver, is used for precise positioning of the sample. The entire system is controlled by an Arduino microcontroller, which handles low-level hardware interfacing and timing operations. A Python-based graphical user interface (GUI) provides high-level control and enables automation of the experimental procedure, including parameter input, stage movement, and data acquisition. Real-time monitoring during movement and ablation is enabled by a webcam positioned in front of the 100~mm CF viewport, supporting accurate experimental setup alignment. Motion commands generated by the GUI (Fig.~\ref{fig:GUI}) are transmitted to the Arduino UNO via serial communication, enabling automatic positioning of the sample before each laser shot. The user specifies the number of measurement locations, translation step size, and scanning sequence through the GUI. After completion of the prescribed number of pulses at one location, the controller automatically translates the sample to a fresh region, thereby preventing overlap between successive ablation craters. Manual positioning is also available to facilitate alignment and sample loading.

	\subsection{Laser synchronization and shutter control}
	
\begin{figure}[hbtp]
		\begin{minipage}{0.45\textwidth}
			\centering
			\includegraphics[width=1\textwidth]{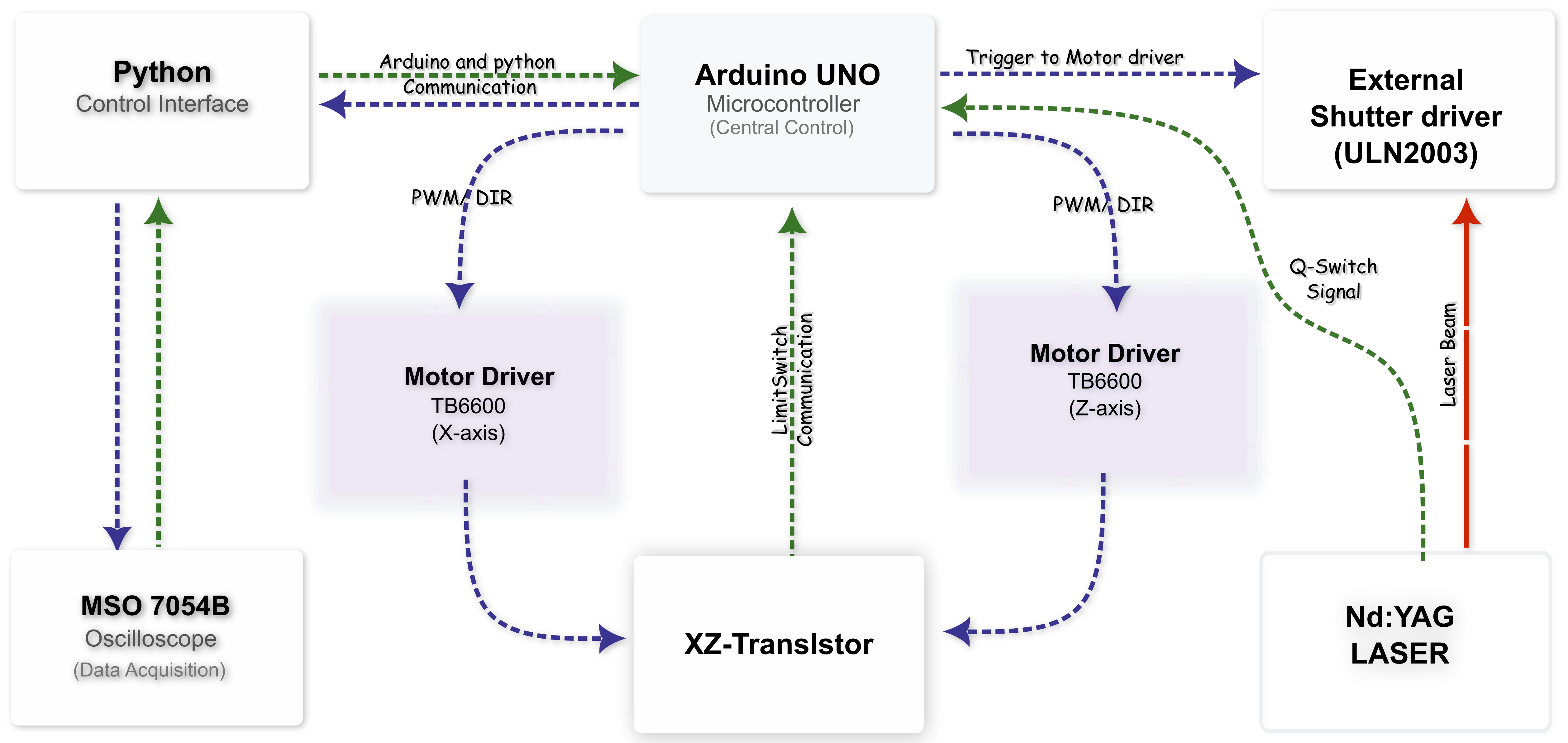}
			\caption{Block diagram of the experimental control and synchronization system. Arduino UNO controls X--Z motion via TB6600 drivers and triggers the shutter (ULN2003) synchronized with the Nd:YAG laser. Data acquisition is performed using an oscilloscope.}
			\label{fig:Exp_sys_block_diagram}
		\end{minipage}
	\end{figure}

\begin{figure}[hbtp]
		\begin{minipage}{0.45\textwidth}
			\centering
			\includegraphics[width=1\textwidth]{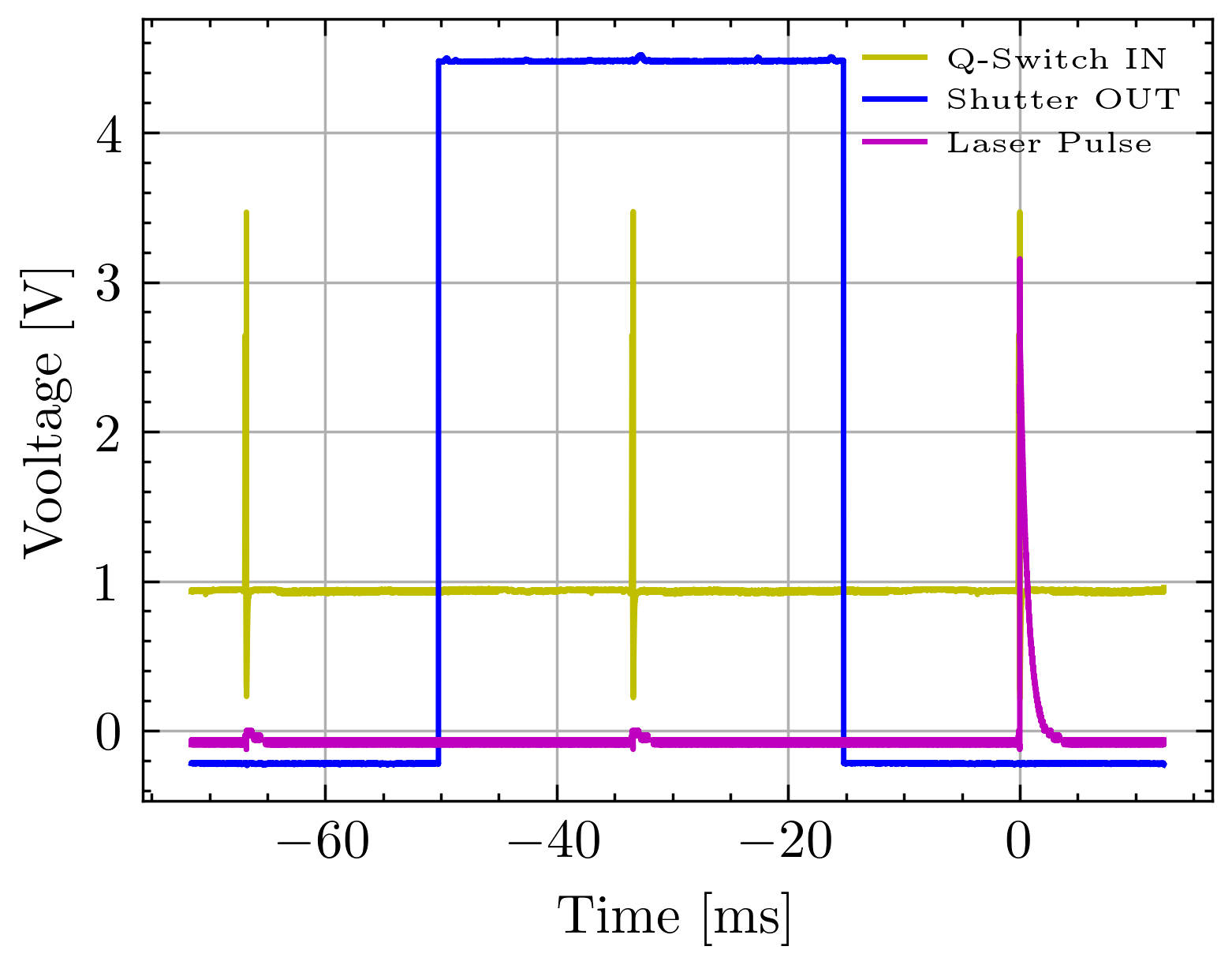}
			\caption{Triggering system of the automated control showing Q-switch signal, 17~ms delay, 35~ms shutter pulse, and 15~ms post-pulse isolation delay.}
			\label{fig:Trigger_dia}
		\end{minipage}
	\end{figure}

	Since the Q-switched Nd:YAG laser operates continuously at a repetition rate of 30~Hz, a rotary solenoid (MSM GDAY 050 X20 E22) is employed for single-pulse selection. The synchronization scheme is illustrated in Fig.~\ref{fig:Exp_sys_block_diagram}, while the detailed timing sequence is presented in Fig.~\ref{fig:Trigger_dia}. 
		
	The 5~V, 30~$\mu$s Q-switch synchronization signal from the laser serves as the master timing reference for the automation system. Upon receiving this trigger, the Arduino UNO generates a precisely timed control pulse of approximately 35~ms duration to actuate the shutter through a ULN2003AN driver module, which supplies the current required for the solenoid. The timing sequence is as follows: after detection of the Q-switch signal, there is a delay of approximately 17~ms, followed by a shutter pulse of 35~ms duration. After the shutter closes, an additional delay of approximately 15~ms ensures proper isolation of the pump laser pulse before the next pulse in the sequence. This shutter pulse is applied to the ULN2003AN driver module with multiple input and output channels connected in parallel to deliver approximately 1~A current required for driving the shutter motor (Fig.~\ref{fig:Exp_sys_connection_diagram}). The timing sequence ensures that only the selected laser pulse reaches the sample while blocking the subsequent pulses, thus enabling reliable single-pulse operation throughout the experiment.

	\subsection{Detector synchronization and data acquisition}\label{sec:Data collection}
	
	The synchronization architecture used for signal acquisition is illustrated in Fig.~\ref{fig:Exp_sys_block_diagram}. The Arduino UNO serves as the central hardware controller, coordinating the motorized translation stage, external shutter driver (ULN2003), and data acquisition system. Experimental parameters entered through the Python-based graphical user interface (GUI) are transmitted to the Arduino via serial communication, which subsequently executes the programmed measurement sequence.
	
	After positioning the sample at the desired location using the X--Z translation stage, the Arduino generates the control signal to actuate the external shutter, thereby synchronizing laser exposure with the acquisition cycle. The Nd:YAG laser operates using its internal Q-switch timing, while a fast photodiode positioned adjacent to the laser provides the master trigger signal for the digital storage oscilloscope (MSO7054B). This trigger initiates single-shot acquisition of the transient signals recorded simultaneously from the reflected and transmitted probe beams using avalanche photodiodes.
	
	In addition to triggering the oscilloscope, the fast photodiode signal is calibrated against the pump laser pulse energy and serves as shot-to-shot energy monitor. The recorded transient signals are subsequently normalized to compensate for pulse energy fluctuations between successive laser shots. Upon completion of each acquisition, the oscilloscope waveform is automatically transferred to the computer through the Python control software, where it is archived together with the corresponding experimental metadata, including sample position, laser energy, ambient pressure, pulse number. This fully automated synchronization sequence minimizes operator intervention and ensures repeatable timing and consistent data acquisition throughout the experiment.

	\subsection{Automated data management}
	
	The data acquisition and processing workflow is coordinated through the GUI shown in Fig.~\ref{fig:GUI}. Prior to each measurement, the user specifies experimental parameters e.g. sample material, film thickness, laser wavelength, chamber pressure, pulse energy, oscilloscope time window, measurement position, and number of pulses. 
	
	The experimental procedure is executed in three sequential stages. In the first stage, laser ablation is carried out at a selected position to remove the thin film from the substrate, typically requiring up to 15 pulses. The reflected and transmitted probe signals are collected as explained in section \ref{sec:Data collection} .After completion of the ablation at a given location, the sample is translated to a fresh region using the motorized X--Z translation stage. The displacement values are specified through the graphical user interface to prevent overlap between successive laser irradiation sites, thereby ensuring independent measurements. The automated acquisition sequence is repeated over multiple locations under identical experimental conditions to evaluate measurement repeatability and improve statistical reliability. In addition to automated motion, manual positioning of both the X and Z axes is available through the GUI to facilitate the sample alignment before and after automated measurements.
	
	Following each laser shot, the acquired waveforms are automatically transferred from the oscilloscope to the host computer and stored in the CSV format using a standardized file-naming convention. Each filename uniquely encodes the corresponding experimental conditions, including the sample material and film thickness, ablation geometry, pump laser wavelength, chamber pressure, pump laser energy, oscilloscope time scale, ablation position, and laser pulse number. This automated naming scheme eliminates manual data entry, ensures complete traceability of every measurement, and enables efficient post-processing and retrieval of experimental datasets. A typical filename has of the form \\
	 \textit{a2\_fp\_1064nm\_1000mbar\_020mj\_010us\_pos1\_p1}.

	The graphical user interface allows the user to define the number of shots, control the stage motion sequence, and manage file storage. As a result, large datasets can be acquired systematically with minimal operator involvement. The integration of shutter control within the automated sequence further ensures safe and controlled operation of the laser system.
	
	\subsection{Data analysis algorithm}
	Following acquisition, the recorded waveforms are automatically processed using dedicated Python software developed for the present work. The processing pipeline begins with shot-to-shot normalization of each waveform using the pump laser energy measured by the fast photodiode, thereby compensating for the pulse energy fluctuation. A baseline correction is subsequently applied to remove any DC offset from the detector signals before further analysis.
	
	The corrected waveforms are then converted into normalized transmission and reflection percentages. Since measurements are performed at multiple locations on the sample, slight variations in the initial reflected signal (typically within 5--15\%) are observed due to local surface non-uniformity, minor alignment differences, and mechanical vibrations of the translation stage. To ensure data quality, the software calculates the standard deviation of the reflected voltage at each measurement location. 
	
	The normalization procedure employs the first and last laser pulses as reference states corresponding to the intact and completely cleaned film, respectively. For the first pulse, the average transmission over the initial 10\% of the recorded waveform is assigned as the 0\% transmission reference or minimum transmission, while the corresponding reflected signal is defined as the 100\% reflection reference or maximum reflection. Conversely, for the final pulse, the average transmission over the final 10\% of the waveform is assigned as the 100\% transmission reference, representing complete film removal, and the corresponding reflection is taken as 0\%. The transmission and reflection signals obtained for all intermediate pulses are subsequently normalized with respect to these reference levels.
	
	To improve statistical reliability, identical experiments are repeated at multiple sample positions (fresh locations) under identical operating conditions as mentioned. After the normalization, the transmission and reflection data for a particular experiment are averaged to obtain the representative transient response. The corresponding standard deviation is simultaneously estimated and stored as the experimental uncertainty. The processed datasets are automatically exported as position-averaged (NAV) files, which constitute the primary input for subsequent analysis. The complete processing workflow is implemented in Python and is publicly available as open-source software for reproducibility~\cite{renjith_software}.
	
	The generated NAV files are subsequently analysed using a dedicated Python-based visualization program that provides automated comparison of transient transmission and reflection measurements acquired under different experimental conditions, including variations in material, film thickness, ablation geometry, ambient pressure, and pump laser energy. The software supports rapid visualization of multiple datasets, enabling efficient comparison and interpretation of large experimental campaigns.
	
	The developed automation framework substantially reduces operator intervention by integrating instrument control, synchronized data acquisition, automatic file management, data processing, and statistical analysis within a unified software environment. The modular architecture readily accommodates future extensions, including closed-loop pressure regulation and automated laser energy stabilization. Consequently, the developed system provides a robust semi-automated experimental platform that improves measurement repeatability, enhances data integrity through standardized processing, and significantly reduces the time required for time-resolved laser ablation experiments.

	\subsection{System Validation and Performance Assessment}
	The developed experimental system was validated to assess the operational reliability of chamber, the motorized sample translation system, the shutter-based exposure sequence, and the automated data-acquisition workflow. Since the primary objective of this work is instrumentation development, the validation emphasized on the repeatability, stability, and ease of operation rather than the exploration of the ablation physics.
	
	The first stage of validation focused on the mechanical performance of the newly developed chamber and sample-handling system. The stainless-steel chamber offered improved structural rigidity and better access to the ports. Its vacuum integrity was evaluated by a helium leak test at an internal pressure of 0.03~mbar, which yielded a leak rate of \(2 \times 10^{-7}\,\mathrm{mbar\,L\,s^{-1}}\), while the background leak rate was \(4 \times 10^{-8}\,\mathrm{mbar\,L\,s^{-1}}\). The motorized translation stage was then repeatedly tested, using camera output and stage-command feedback, to confirm reliable repositioning of the sample to fresh locations during the experiment as well as during manual operation. This capability is essential for preventing overlap between successive ablation events and for ensuring uniform exposure of the thin-film target.
	
	The second stage of validation addressed the optical and timing sequence. The shutter operation was verified to ensure reliable opening and closing throughout the experimental cycle. In addition, the sequence of sample motion, shutter actuation, laser exposure, trigger generation, and signal acquisition was confirmed to be properly synchronized through the control interface. The use of a fast photodiode as the trigger reference ensured that oscilloscope acquisition remained temporally aligned with the arrival of the pump laser pulse.
	
	The third stage of validation involved the automated data-acquisition framework. A Python-based interface was tested over multiple experimental cycles, and the acquired waveforms were stored automatically with appropriate shot numbering and file organization in folders named according to the date of the experiment. This reduced manual intervention and minimized the likelihood of data loss or labeling errors during long measurement runs. The automation was also assessed for its ability to support repeated measurements over a range of target positions without requiring continuous operator supervision.
	
	Overall, the validation demonstrates that the developed system functions as a reliable semi-automated platform for laser ablation experiments. Although pressure control and laser-energy regulation are not yet fully automated, the present setup successfully automates the most repetitive and error-prone steps in the workflow, thereby improving the repeatability, efficiency, and experimental throughput.

	\section{Results and Discussion}

	The developed laser ablation platform was evaluated primarily from an instrumentation perspective, with emphasis on mechanical robustness, operational flexibility, synchronization, and automation of the measurement workflow. The newly developed stainless-steel chamber, motorized target translation system, shutter control, and automated data-acquisition framework together constitute a practical experimental platform for time-resolved laser ablation studies. One of the major outcomes of the development is the ability to perform repeated measurements with minimal manual intervention, consistent sample positioning, and systematic organization of the acquired data.

	The automated experimental system completes a single measurement cycle in approximately 37~s, including data acquisition and storage for 15 laser pulses at one ablation position. Each pulse generates nearly $2\times10^5$ data points per channel across four channels, corresponding to a temporal window of up to 100~$\mu$s after ablation. A complete experimental set involving three positions under identical conditions is completed in approximately 112~s, with a total of 45 data files recorded. The dominant contribution to the total acquisition time arises from data transfer from the oscilloscope to the PC. In comparison, manual operation requires nearly 1~min to complete a single experiment and save a single data file.

	\begin{figure}[htbp]
		\centering
		
		\begin{subfigure}[t]{0.48\columnwidth}
			\centering
			\includegraphics[width=\linewidth]{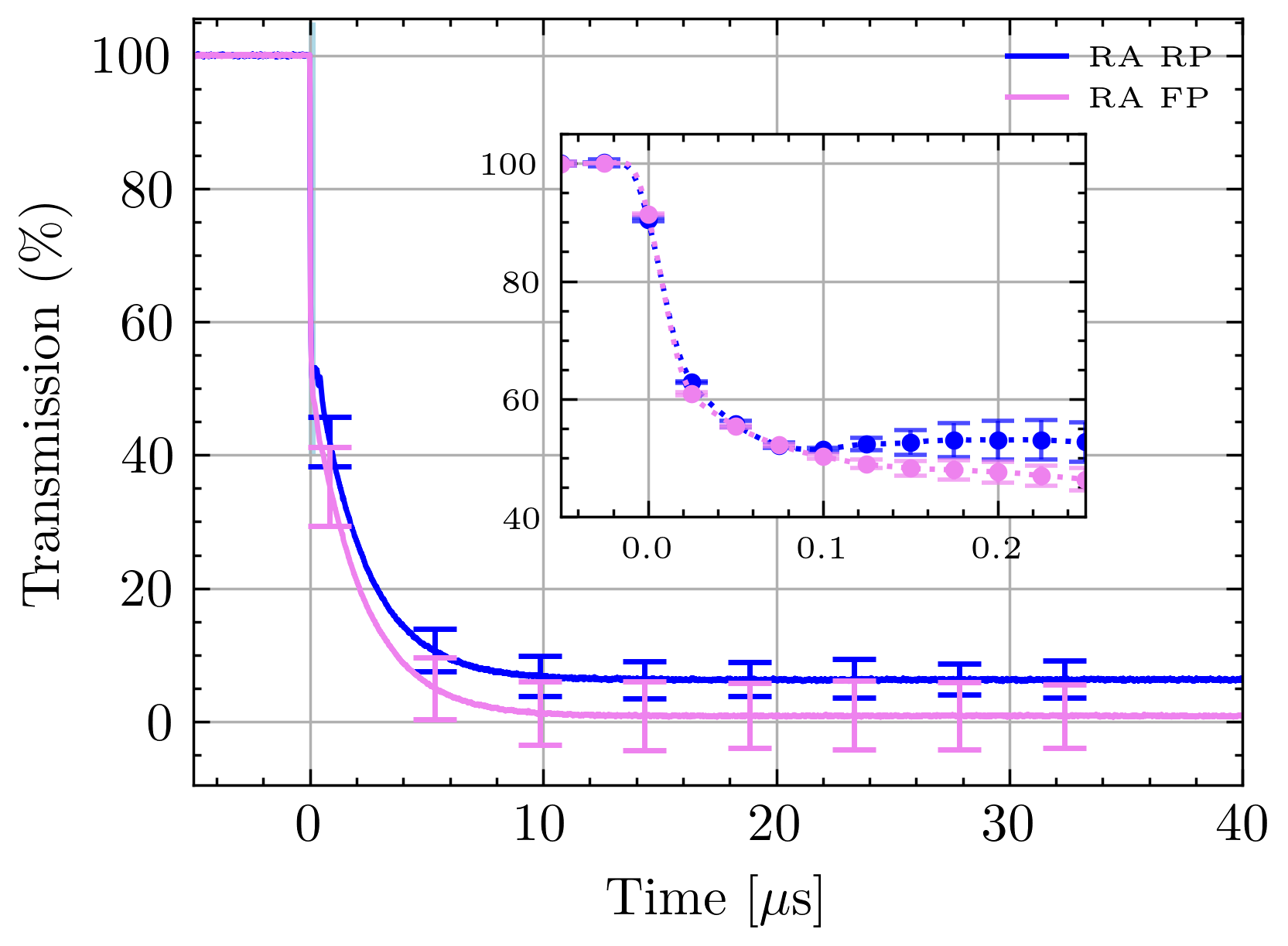}
			\caption{Rear ablation reflection for front vs rear probing. The inset is a magnified portion up to 100ns of the reflection with data points plotted at 10-point intervals.}
			\label{fig:a4_ra_15j_005e-1mbar_p1_refle}
		\end{subfigure}
		\hfill
		\begin{subfigure}[t]{0.48\columnwidth}
			\centering
			\includegraphics[width=\linewidth]{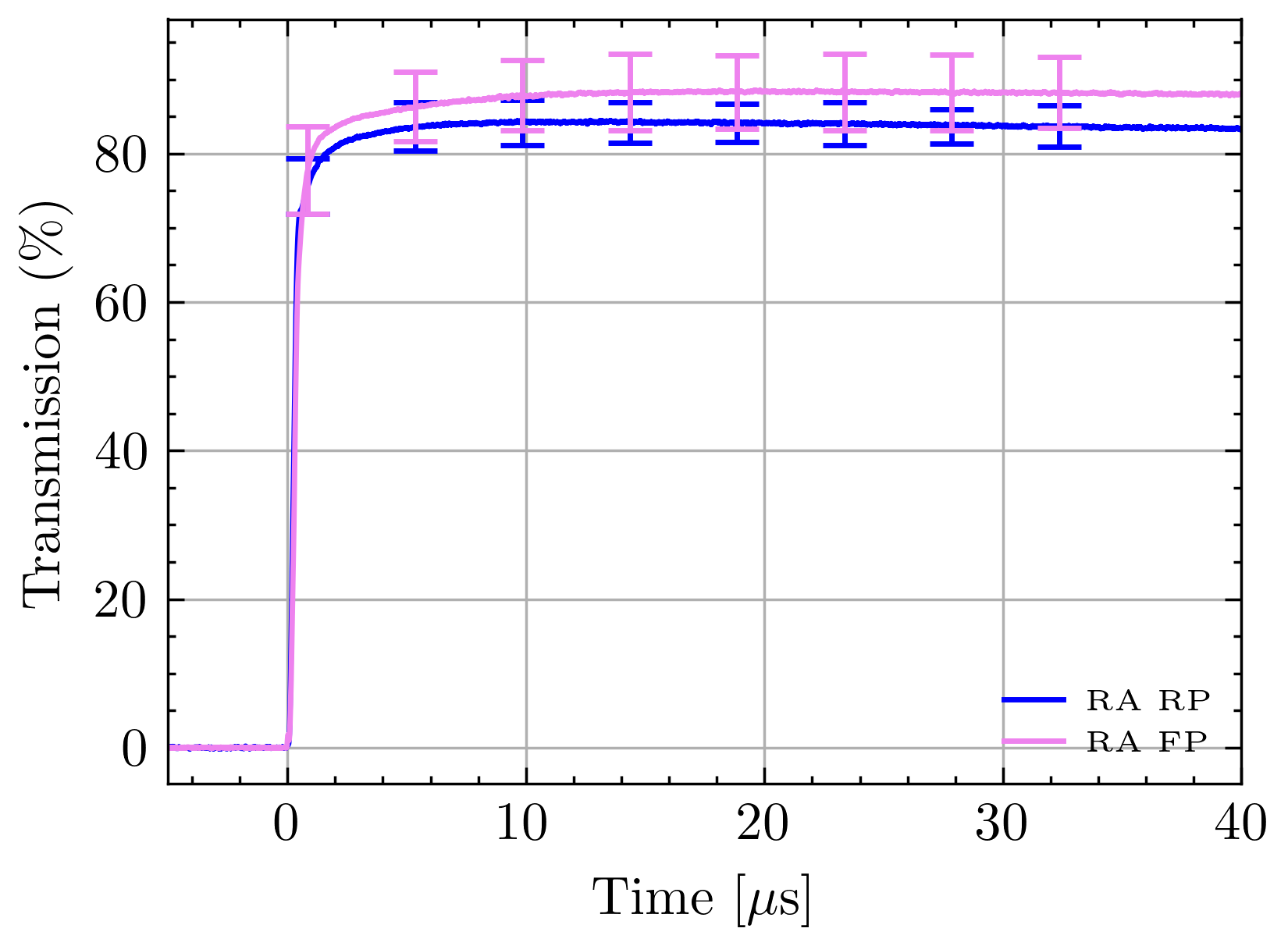}
			\caption{Rear ablation transmission for front vs rear probing}
			\label{fig:a4_ra_15j_005e-1mbar_p1_trans}
		\end{subfigure}
		
		\vspace{0.3cm}
		
		\begin{subfigure}[t]{0.48\columnwidth}
			\centering
			\includegraphics[width=\linewidth]{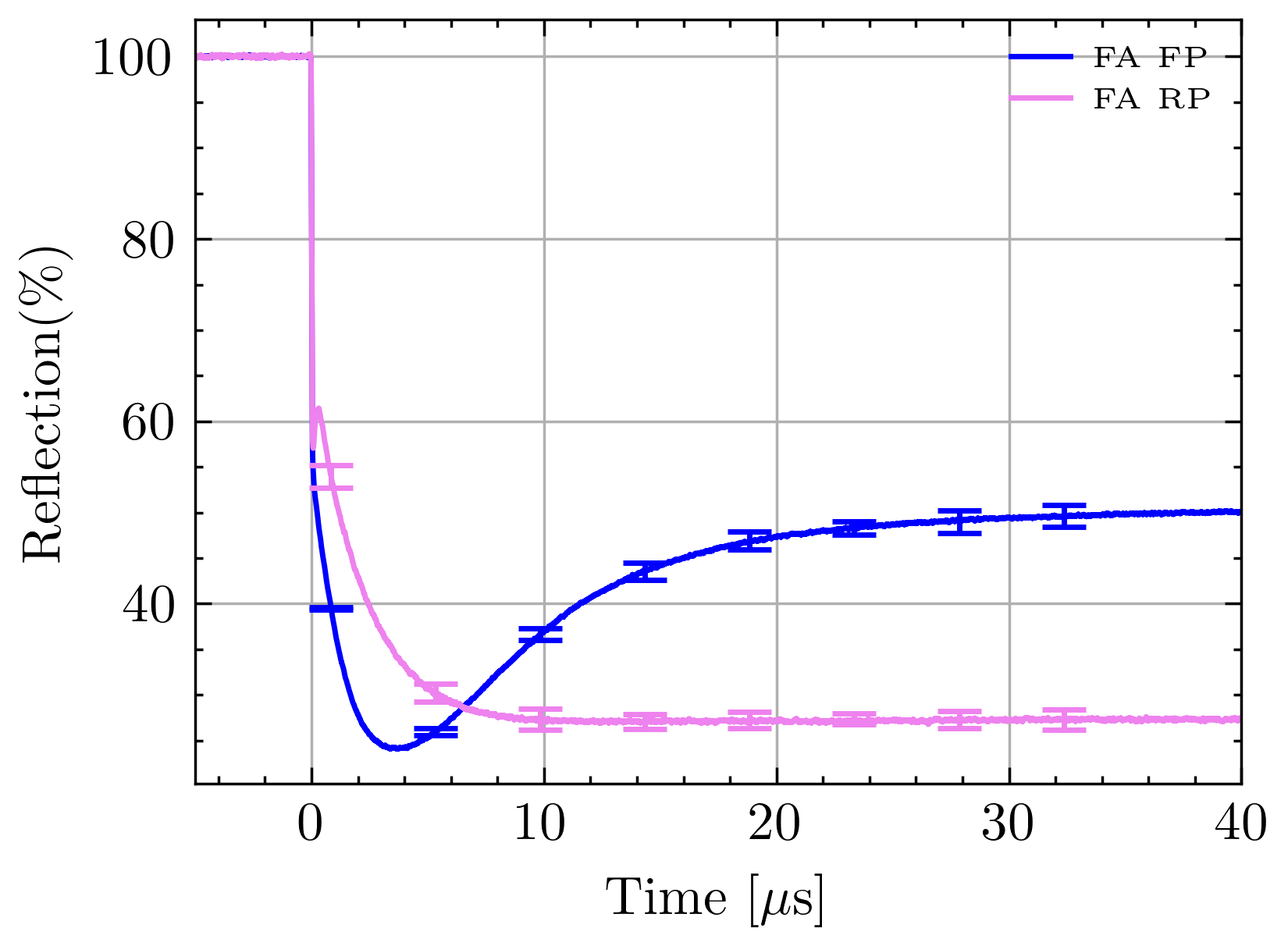}
			\caption{Front ablation reflection for front vs rear probing}
			\label{fig:a4_fa_15j_005e-1mbar_p1_refle}
		\end{subfigure}
		\hfill
		\begin{subfigure}[t]{0.48\columnwidth}
			\centering
			\includegraphics[width=\linewidth]{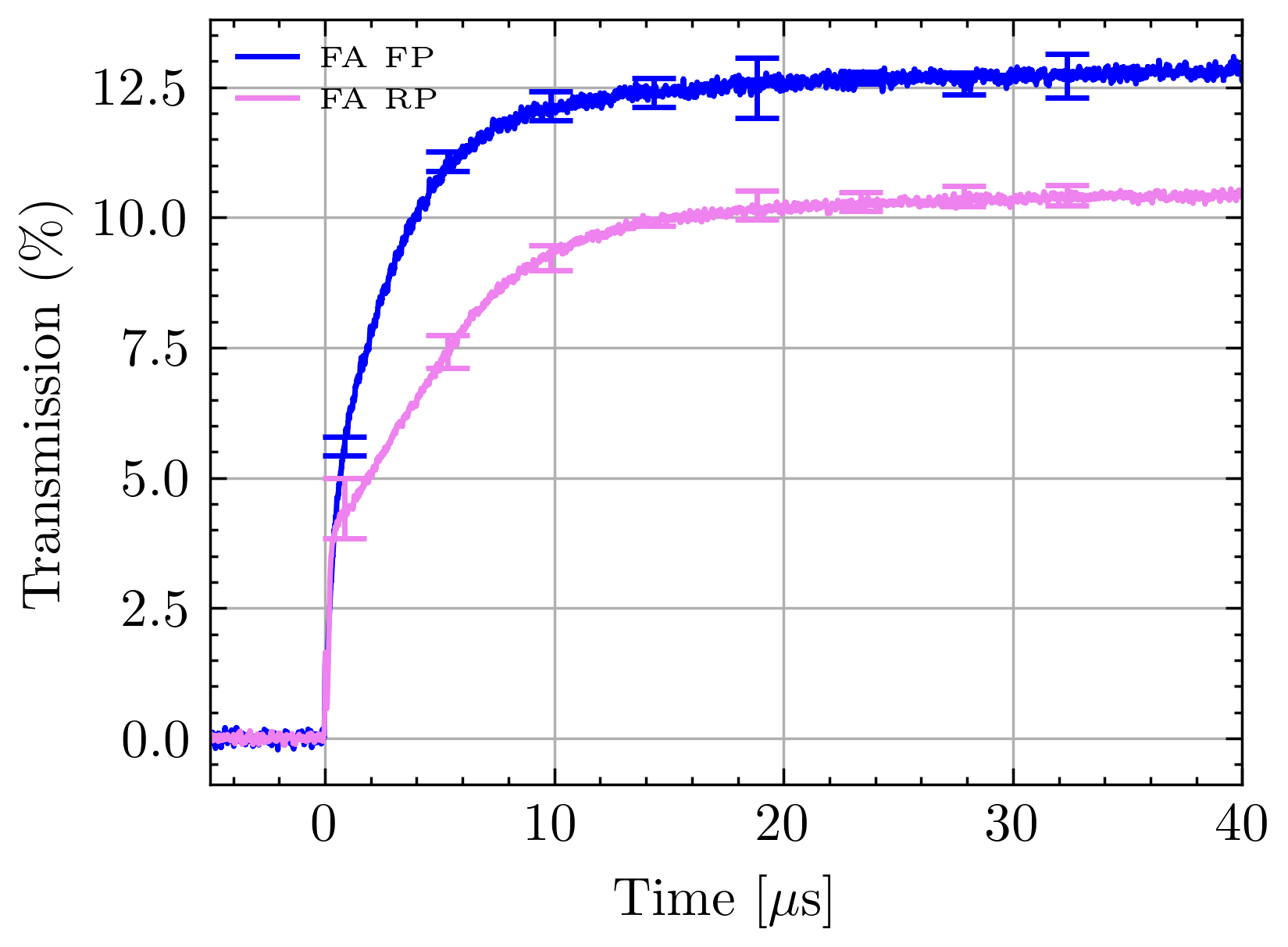}
			\caption{Front ablation transmission for front vs rear probing}
			\label{fig:a4_fa_15j_005e-1mbar_p1_trans}
		\end{subfigure}
		
		\caption{Typical time-resolved reflection and transmission probe signals acquired using the developed pump--probe diagnostic system for a 400~nm aluminum film at a pump fluence of 15.3~J/cm$^2$ and a background pressure of 0.5~mbar. The figure compares the transient optical responses obtained for different laser ablation (front and rear) and probe (front and rear) geometries.}
		\label{fig:a4_ra_fa_15j_005e-1mbar}
	\end{figure}

	Typical time-resolved reflected and transmitted probe signals measured under front-side ablation with rear-side probing of a 400~nm thick aluminum film coated on quartz substrate of 50x50x4mm substrate at a pump fluence of 15.3~J/cm$^2$ and a background pressure of 0.5~mbar are shown in Fig.~\ref{fig:a4_ra_fa_15j_005e-1mbar}. To enhance the visibility of data, the error bars are kept only for a few instances. Prior to the arrival of the pump pulse ($t<0$), the reflection signal remains stable at approximately 100\%, while the transmission is negligible, indicating proper alignment and steady baseline conditions of the detection system. 
	
	At $t \approx 0$, corresponding to the arrival of the pump pulse, a sharp decrease in the reflected signal is observed along with a simultaneous rise in the transmitted signal. This rapid transition confirms the successful alignment between the pump laser, probe beam, and data-acquisition system. The smooth temporal evolution of both signals demonstrates the capability of the setup to capture transient changes with adequate temporal resolution.
	
	A comparison of the different probing geometries reveals several interesting physical phenomena. Figures~\ref{fig:a4_ra_15j_005e-1mbar_p1_refle} and \ref{fig:a4_ra_15j_005e-1mbar_p1_trans} compare the reflection and transmission signals obtained using front- and rear-side probing during rear ablation. It is evident that both probing configurations exhibit nearly identical temporal behavior, indicating that the transient optical response is largely independent of the probe direction under rear-ablation conditions.
	
	In contrast, a significant difference is observed for front ablation. Figures~\ref{fig:a4_fa_15j_005e-1mbar_p1_refle} and \ref{fig:a4_fa_15j_005e-1mbar_p1_trans} show that the transient signals strongly depend on the probing geometry. In particular, the reflection signal measured with front probing (FA--FP) exhibits a pronounced dip, whereas this feature is absent in the rear probing (FA--RP) measurement. This suggests that the transient processes occurring at the irradiated surface differ considerably from those observed through the substrate, highlighting the sensitivity of the measurement towards the probe geometry~\cite{renjith_fa_p1}. A similar geometry-dependent variation is also observed in the transmission. For a fixed ablation geometry, the transient reflectivity and transmission response depends strongly on the probing side, and this dependence itself carries information about the ablation dynamics that a single probing geometry cannot provide.

	\begin{figure}[htbp]
		\centering
		
		\begin{subfigure}[t]{0.48\columnwidth}
			\centering
			\includegraphics[width=\linewidth]{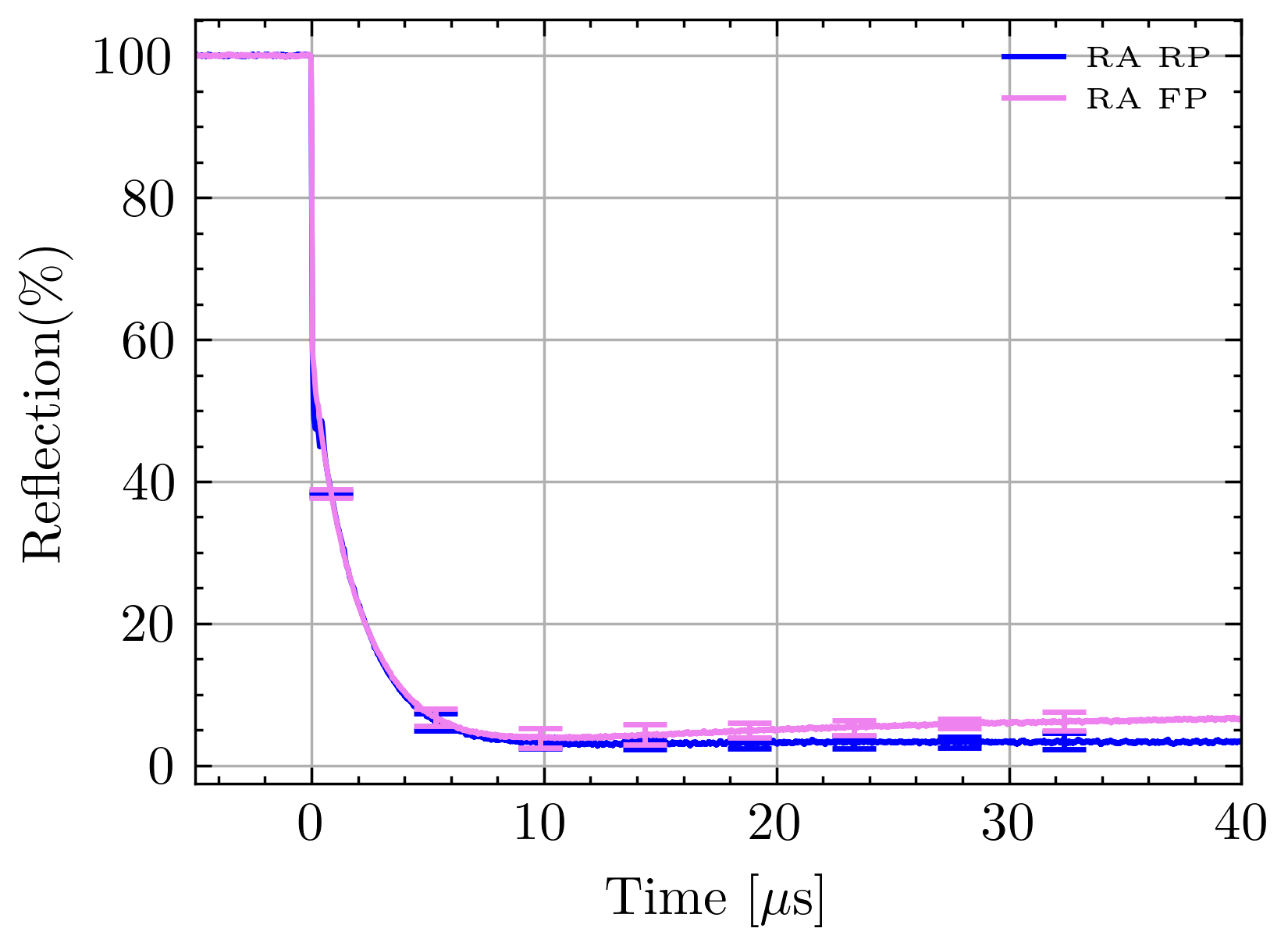}
			\caption{Rear ablation reflection for front vs rear probing}
			\label{fig:s4_ra_25j_005e-1mbar_p1_refle}
		\end{subfigure}
		\hfill
		\begin{subfigure}[t]{0.48\columnwidth}
			\centering
			\includegraphics[width=\linewidth]{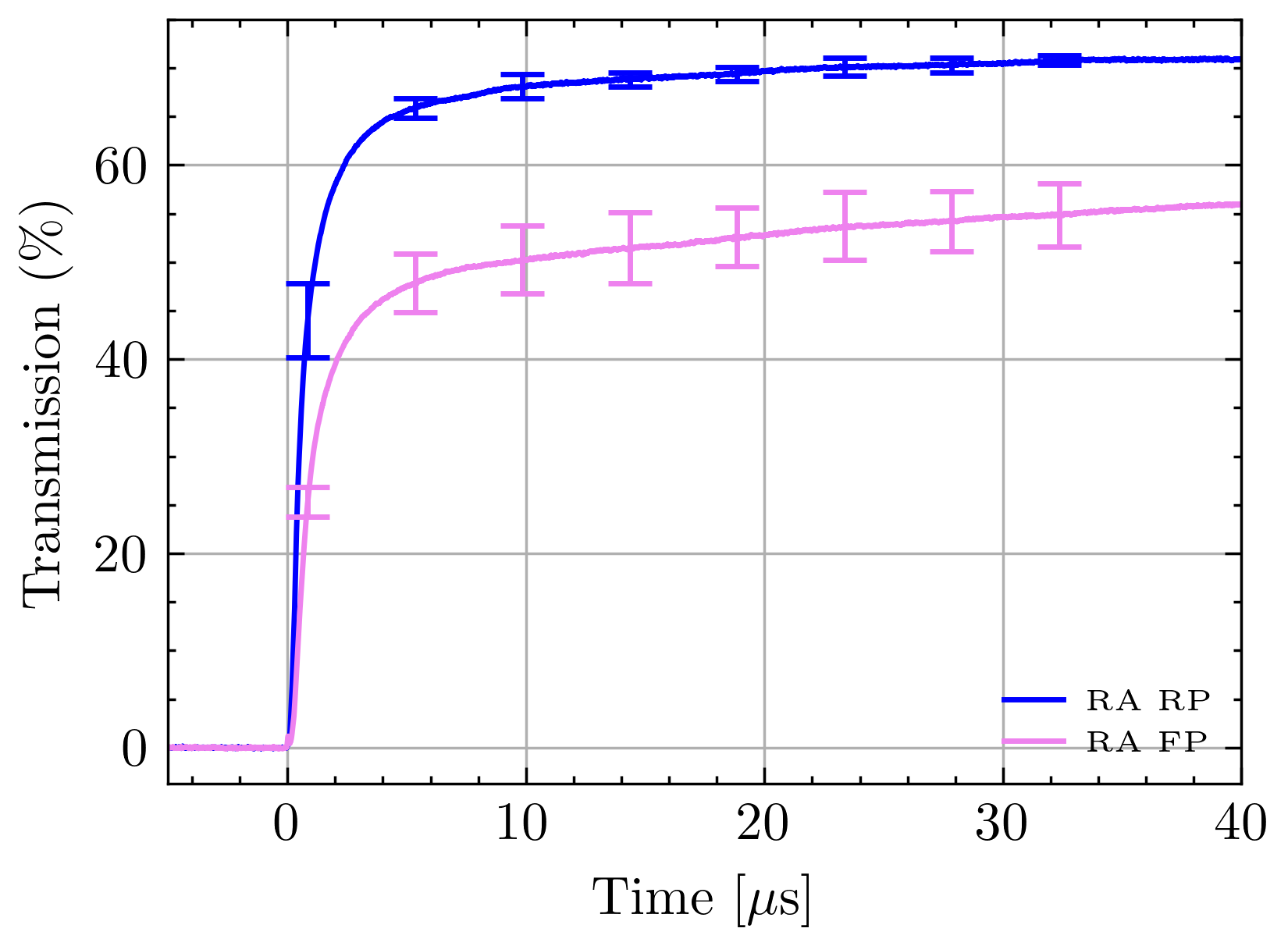}
			\caption{Rear ablation transmission for front vs rear probing}
			\label{fig:s4_ra_25j_005e-1mbar_p1_trans}
		\end{subfigure}
		
		\vspace{0.3cm}
		
		\begin{subfigure}[t]{0.48\columnwidth}
			\centering
			\includegraphics[width=\linewidth]{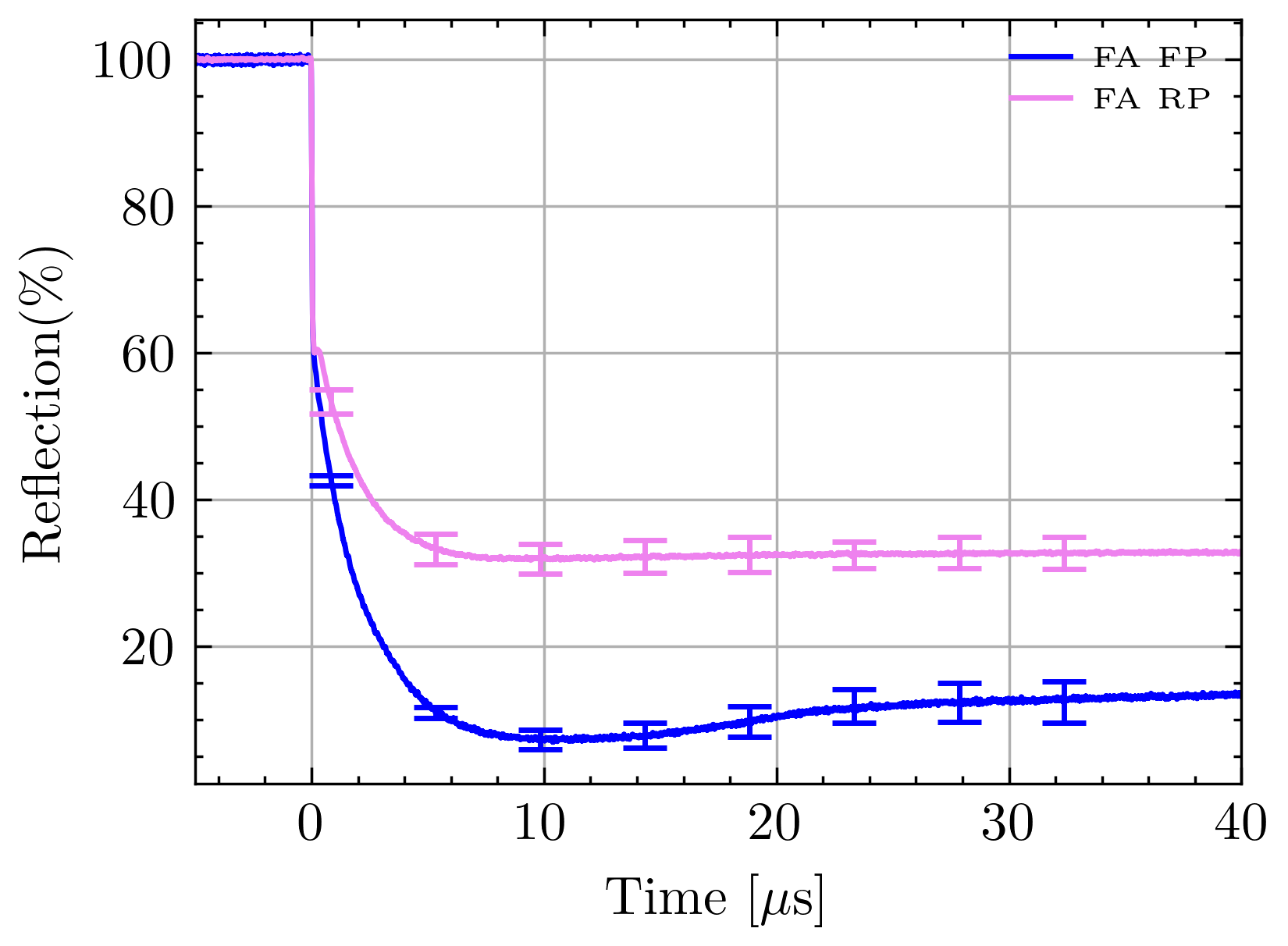}
			\caption{Front ablation reflection for front vs rear probing}
			\label{fig:s4_fa_25j_005e-1mbar_p1_refle}
		\end{subfigure}
		\hfill
		\begin{subfigure}[t]{0.48\columnwidth}
			\centering
			\includegraphics[width=\linewidth]{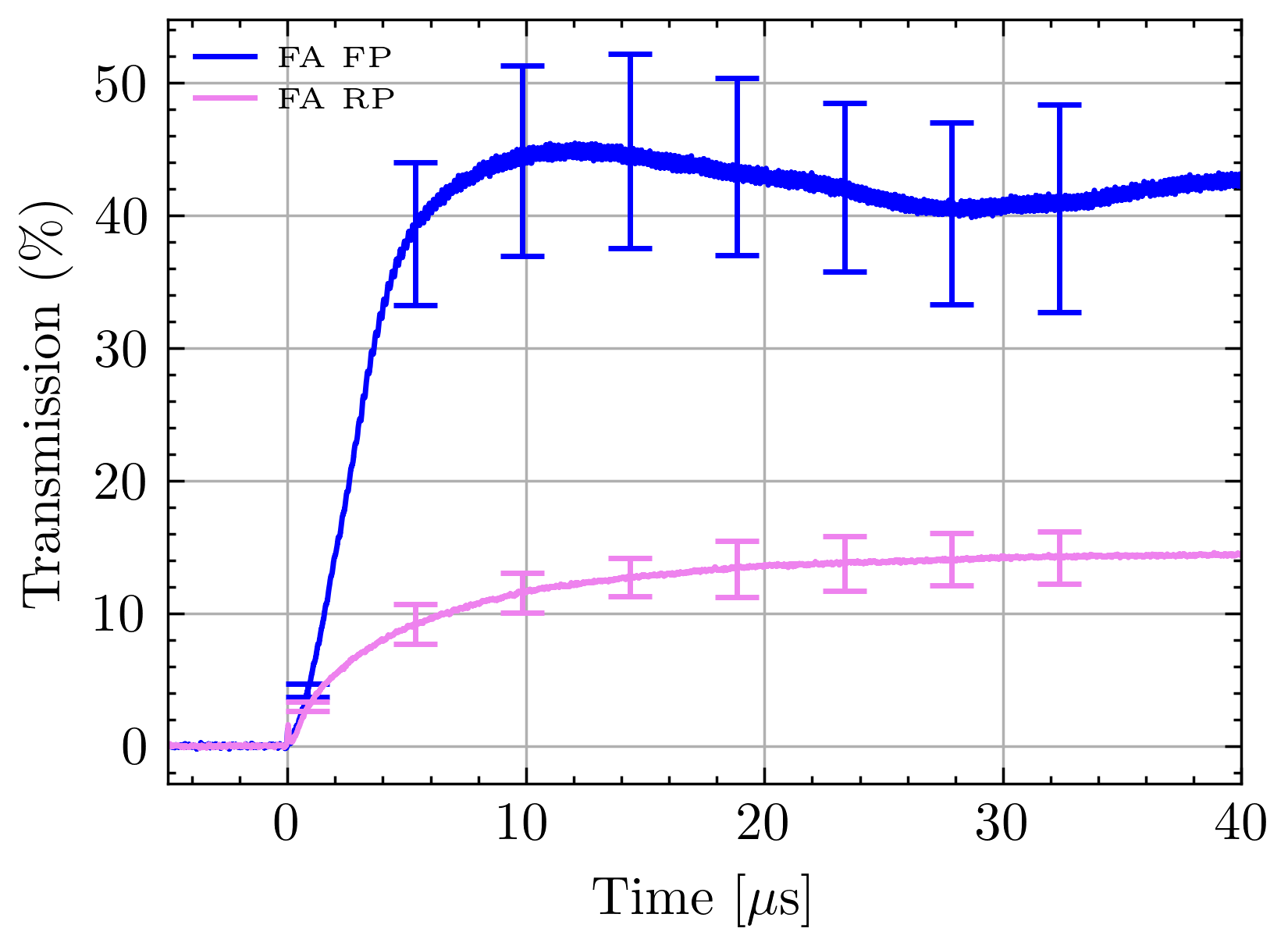}
			\caption{front ablation transmission for front vs rear probing}
			\label{fig:s4_fa_25j_005e-1mbar_p1_trans}
		\end{subfigure}
		
		\caption{Typical time-resolved reflection and transmission probe signals acquired using the developed pump--probe diagnostic system for a 400~nm silver film at a pump fluence of 25.0~J/cm$^2$ and a background pressure of 0.5~mbar. The figure compares the transient optical responses obtained for different laser ablation (front and rear) and probe (front and rear) geometries. }
		\label{fig:s4_ra_fa_25j_005e-1mbar}
	\end{figure}

	The same experiment was performed on a 400~nm silver thin film coated on quartz substrate of 50x50x4mm substrate to investigate whether the observed geometry-dependent response is material specific. Figures~\ref{fig:s4_ra_25j_005e-1mbar_p1_refle} and \ref{fig:s4_ra_25j_005e-1mbar_p1_trans} compare the reflection and transmission signals obtained using front- and rear-side probing during rear ablation. Similar to aluminum, both probing configurations exhibit nearly identical temporal behavior in the reflection signal, while only a small difference is observed in the transmission response.
	
	In contrast, a pronounced dependence on the probing geometry is observed during front ablation, as shown in Figures~\ref{fig:s4_fa_25j_005e-1mbar_p1_refle} and \ref{fig:s4_fa_25j_005e-1mbar_p1_trans}. The reflection signal measured using front-side probing (FA--FP) decreases rapidly to nearly zero, whereas the rear-side probing (FA--RP) signal remains at a significantly higher level throughout the measurement. A similar trend is observed in the transmission signals, where the front-side probing configuration records substantially higher transmission than the rear-side probing configuration. These observations indicate that the transient optical properties differ significantly at the irradiated surface and through the substrate, emphasizing the importance of employing multiple probing geometries to fully characterize the laser-induced ablation process.
	
	Compared with aluminum, the differences between the probing geometries are more pronounced in silver. This can be attributed to the higher reflectivity and larger latent heat of vaporization of silver, which modify the energy deposition and subsequent material removal dynamics. Consequently, the optical response measured from the irradiated surface is likely to differ more significantly from that observed through the substrate, leading to an inference of stronger dependence on the probe geometry.

	The present work, therefore, demonstrates that a carefully designed experimental chamber with control architecture can significantly enhance experimental reliability, efficiency, stability, and reproducibility.
	
	\section{Conclusion}

	A modular experimental platform for time-resolved investigation of laser ablation and plume propagation in a thin films has been successfully developed and validated. The system provides a stable, repeatable, and flexible experimental environment for pump–probe studies by integrating a stainless-steel vacuum chamber, motorized sample positioning, a shutter-based single-pulse selection scheme, and an Arduino–Python-based automation architecture. The modular design enables both front-side and rear-side ablation geometries without requiring major modifications to the experimental configuration, thereby improving experimental flexibility and efficiency.

	The developed platform overcomes several limitations associated with conventional time-resolved laser ablation experiments, In addition, the results demonstrate that probing direction serves as an independent diagnostic variable. for a fixed ablation geometry, the transient response measured from the irradiated surface can differ substantially from that measured through the substrate, a finding that could not have been obtained from a single probing geometry. Pump–probe measurements performed on different thin-film samples demonstrate accurate synchronization in which the temporal evolution of the ablation process is reconstructed from multiple laser-ablation shots acquired at different delay times, thereby relying on the assumption of shot-to-shot reproducibility. In contrast, the present system enables reliable single-pulse measurements with precise temporal synchronization, controlled sample positioning, and integrated data acquisition. Pump–probe measurements performed on different thin-film samples demonstrate accurate synchronization, stable optical alignment, and consistent transient reflectivity and transmission signals, validating the platform as an effective diagnostic tool for investigating laser–matter interaction, ablation dynamics, and laser-produced plasma evolution with a temporal resolution better than 15 ns.
		
	Beyond fundamental studies, the versatility of the setup makes it suitable for a broad range of laser processing applications, including laser cleaning of tokamak viewports and mirrors, pulsed laser deposition (PLD), laser-induced forward transfer (LIFT), and surface structuring. Future integration of features such as automated pressure control, laser energy stabilization, and closed-loop feedback can further enhance the precision and autonomy of the platform. Overall, the developed experimental system provides a robust, extensible, and efficient foundation for advanced time-resolved investigations of laser–matter interaction in both fundamental research as well as technological applications.
	
	\section*{Data availability} 
	
	The data that supports the observations of this study are available from the corresponding author upon reasonable request.
	\section*{Acknowledgment} 
The authors would like to thank the Mechanical Workshop at the Institute for Plasma Research (IPR) for their active involvement in fabricating the components used in the experimental setup.


	\end{document}